# Computing High-Degree Polynomial Gradients in Memory

T. Bhattacharya[1*], G. H. Hutchinson[1], G. Pedretti[2], X. Sheng,[2] J. Ignowski,[2] T. Van Vaerenbergh[3], R. Beausoleil[3], J.P. Strachan[4] & D.B. Strukov[1*]

**Abstract**

Specialized function gradient computing hardware could greatly improve the performance of state-of-the-art optimization algorithms, e.g., based on gradient descent or conjugate gradient methods[1,2], that are at the core of control[3], machine learning[4], and operations research[5] applications. Prior work on such hardware, performed in the context of the Ising Machines[6] and related concepts[7,8], is limited to quadratic polynomials and not scalable to commonly used higher-order functions. Here, we propose a novel approach for massively parallel gradient calculations of high-degree polynomials, which is conducive to efficient mixed-signal in-memory computing circuit implementations[9-12] and whose area complexity scales linearly with the number of variables and terms in the function and, most importantly, independent of its degree. Two flavors of such an approach are proposed. The first is limited to binary-variable polynomials typical in combinatorial optimization problems[5], while the second type is broader at the cost of a more complex periphery. To validate the former approach, we experimentally demonstrated solving a small-scale third-order Boolean satisfiability problem[13] based on integrated metal-oxide memristor crossbar circuits[11], one of the most prospective in-memory computing device technologies, with competitive heuristics algorithm[14]. Simulation results for larger-scale, more practical problems[15] show orders of magnitude improvements in the area, and related advantages in speed and energy efficiency compared to the state-of-the-art. We discuss how our work could enable even higher-performance systems after co-designing algorithms to exploit massively parallel gradient computation.

[1]Department of Electrical and Computer Engineering, University of California at Santa Barbara, Santa Barbara, CA 93106, USA; [2]Artificial Intelligence Research Lab, Hewlett Packard Labs, Milpitas, CA, USA; [3]Large Scale Integrated Photonics, Hewlett Packard Labs, Milpitas, CA, USA; [4]Peter Grünberg Institut (PGI-14) and RWTH Aachen University, Germany. *Correspondence and material requests should be addressed to T.B. and D.B.S. (email: tinish@ucsb.edu, strukov@ece.ucs.edu).

**Introduction**

As the growing demand for computing processing power can be no longer supported by semiconductor technology scaling, more focus is now on developing application- and function-specific hardware accelerators. Computing gradients is essential across many applications, such as training the weights in modern Deep Neural Networks (DNN), or in physics-inspired computing paradigms including Ising machines (IMs)[6,16] or closely related approaches with Hopfield neural networks (HNNs)[7] and Boltzmann machines (BMs)[8]. For example, a continuous-time second order HNN[17] consists of a recurrently connected network of pairwise symmetrically coupled graded-response neurons (Fig. S1). Neuron states are continuously updated to seek the minimum of an associated scalar energy function, i.e., a Hamiltonian function in the context of Ising models[17]. For pair-wise couplings, the resulting energy function is quadratic, and hence IMs/HNNs have been extensively applied to solving quadratic unconstrained binary optimization (QUBO) problems[6] where the minimum of an arbitrary quadratic binary function is sought.

The above neuron dynamics effectively depend on the partial derivatives of the energy function with respect to the neuron values (Fig. S1). Therefore, to rapidly converge, the most promising HNNs/IMs hardware implementations rely on massively parallel computations of gradients[6,18]. Such hardware currently constitutes the state-of-the-art in specialized gradient computing circuits. Especially promising are crossbar-circuit implementations based on analog memory devices[19,20], most importantly very dense memristors[21-24], due to prospects of efficient in-memory computing[9,10] and low footprint multi-bit implementations of the coupling weights. Though there have been proposals of higher-($K$)-order HNNs (see, e.g., Fig. S2) and their variants[25-31] that rely on computing gradients of $K$-degree polynomials, an efficient hardware implementation is lacking, while the previous in-memory computing proposals do not readily extend to larger $K$. For example, a straightforward in-memory crossbar array implementation of $K$-order HNN with $N$ neurons (e.g., $N$ variables in energy function) requires $\sim N^K$ coupling weights (Fig. S2), i.e., hardly practical for larger $N$ and/or $K$.

Meanwhile, functions of higher $K$ can represent increasingly important and challenging problems. For example, polynomial unconstrained binary optimization (PUBO) problems[32] are described by $K$-degree polynomials and naturally arise in protein folding and other first-principle calculation methods[33-36] and operations research[37-39]. Notably, $K$ grows linearly with the size of molecular system (molecular orbitals) in PUBO approaches for calculating electronic structure[36]. The well-known $K$–Boolean–satisfiability ($K$-SAT) problem goes from polynomial to expected exponential runtime as $K$ increases from 2 to 3[40]. While higher $K>3$ can be mapped to the $K=3$ case, this requires a polynomial increase in the variables, potentially increasing the runtime with increasing $K$. Interestingly, the original Hopfield network with quadratic ($K=2$) energy function has been extended to much higher memory capacities by utilizing higher-order ($K>2$) energy functionals[41]. However, the operation of such networks, as well as other artificial neural network with high order synapses,[26,42,43] relies on efficient computations of higher order polynomial gradients.

A main contribution of this work is the development of a paradigm for computing gradients of arbitrary-degree polynomial functions in a massively parallel fashion. The proposed paradigm enables efficient hardware that can immediately impact the above described use cases and can be more broadly applied to accelerate gradient computation of arbitrary functions when using Taylor expansion approximation.

**Parallel Gradient Computation**

We expound upon binary-variable function ($H$) gradient computations using a high degree polynomial consisting of four variables ($x$) and four terms (i.e., monomials) of varying degree, each with unique factor $a$ (Fig. 1a). Computing the gradient of such a function requires calculations of all its (pseudo) partial derivatives $\Delta H_{xi} \equiv H_{xi}\Delta x_i$, where $H_{xi}$ is the difference quotient[44] of $H$ with respect to $x_i$ and $\Delta x_i$ is the change in the variable value (Supplementary Note 1). In our approach, pseudo partial derivatives are decomposed into "make" and "break" components. (This terminology is inspired by stochastic algorithms used for solving SAT problems[45].) The make component of pseudo derivative with respect to a given variable represents the contribution from monomials that would change their value from a zero to a non-zero one, i.e., corresponding $a$ value, after flipping the variable state. Similarly, the break component represents the contribution to the derivative from monomials that were previously evaluated to non-zero value but would become zero after changing the variable state. The difference between the make and break values is related to the pseudo partial derivative as $\Delta H_{xi} = (\text{make}_{xi} - \text{break}_{xi})$ - see an example for a specific variable assignment in Fig. 1a. We call the monomials that contribute to make and break value summations as, correspondingly, make and break monomials. Note that the sets of make and break monomials do not overlap and depend on the current variable assignment and that some monomials might be of neither make nor break type. Our goal is to design efficient hardware that can identify such make, break monomials, and then compute make and break values for each variable.

Before discussing in-memory computing implementation, it is convenient to visualize high-order polynomials as a bipartite graph with variables and monomials representing two sets of vertices (Fig. S3). Such a graph maps naturally to a crossbar memory array by assigning variables to one set of (say, vertical) crossbar wires while monomials to another (horizontal) set of wires by setting the array's binary coupling weight to the "on" state, i.e., $b_{ji} = 1$, if the $i$-th variable is present in a $j$-th monomial, while setting the weight to the "off" state ($b_{ji} = 0$) otherwise (Fig. 1b). All pseudo derivatives are computed in parallel in three steps. First, in a "forward pass", variable values are applied to the crossbar array, and a dot product $\sum_i b_{ji} x_i$ corresponding to the sums of monomial variables are calculated at the monomial side of the crossbar array (Fig. 1c). This information is then used to identify all make and break monomials (Fig. 1d,e). Specifically, monomials whose variable sum is equal to their maximum value, i.e., monomial degree $K_j$, are break monomials, while those whose sums are one short of their degree ($K_j$-1) are make monomials. For example, for the specific variable assignment of the considered function, there are two (2nd and 3rd degree) make monomials and one (1st degree) break monomial (Fig. 1d,e).

The data flow via the crossbar circuit is reversed in the next two steps. Unit inputs for the identified make monomials, denoted by indicator variable $s_j$, and zero inputs otherwise, are applied at the monomial side of the array to compute the number of make monomials that each variable is a member of at the variable side of the array. In a more general case shown in Fig. 1d, in a "backward" make pass, unit inputs are scaled according to the monomial factors so that the computed dot products $\sum_j b_{ji} a_j s_j$, called make variable sums, correspond exactly to variables' potential make values. "Potential" because computed values are only relevant for currently zero variables, and only flipping those variables can change a monomial value to a non-zero one. Therefore, the make values are computed by multiplying the make variable sums at the crossbar array periphery by the inverted value of a variable, i.e., $\bar{x}_i \sum_j b_{ji} a_j s_j$, thus making the result zero for all variables that are currently one. Similar operations are performed in a "backward" break pass to compute break values, with the only difference that scaled unit inputs are applied according to the identified break monomials, denoted by indicator variable $z_j$, and the final peripheral multiplication is performed with normal variable values to compute the break values $x_i \sum_j b_{ji} a_j z_j$ (Fig. 1e). Supplementary Note 2 provides a more formal, rigorous framework for the proposed in-memory massively parallel computation of function pseudo gradient.

There are several important variations and extensions of the discussed approach. Monomial variable sums can be compared to fixed values (0 and 1) to identify make and break monomials by applying inverted values of variables in the forward pass (Fig. S4a-c). This allows for simplifying peripheral circuitry of the backward passes by not requiring storing specific thresholds for each monomial of Fig. 1d,e approach. An input scaling in the backward passes can be implemented with more complex analog (multi-level) memory devices to simplify the periphery further (Fig. S4d). Also, backward pass computations can be implemented on separate crossbar arrays, which allows performing all three operations in parallel in a pipelined design to increase computational throughput (Fig. S4e).

More importantly for further discussion, the proposed approach is also suitable after minor modifications for computing pseudo gradients of Boolean logic functions expressed in conjunctive normal form (CNF), thus allowing solving Constraint Satisfaction Problems (CSPs) like SAT and MAXSAT, in native space without potentially time-consuming conversion to an equivalent PUBO form (see Supplementary Note 3 for details on such conversion). CNF Boolean function comprises of conjunction (AND) of clauses, where a clause is a disjunction (OR) of literals, i.e., normal or complementary variables. The goal in the $K$-SAT problem, known as NP-hard for $K \geq 3$, is to find a variable assignment that satisfies all clauses of the given CNF function, with up to $K$ literals per clause. Fig. S5 shows details of all steps for parallel computation of Boolean variable gain values. (A gain value is commonly used in SAT community to describe the change in the number of satisfied clauses after flipping a variable[14]. It is effectively the negative of the pseudo partial derivative with respect to that variable of an equivalent PUBO energy function.) In this case, clauses are mapped to crossbar array rows, while literals (Fig. S5b), are mapped to crossbar array columns. Forward (Fig. S5c) and backward (Fig. S5d,e) operations are similar to those of the monomial approach with inverted input variables (Fig. S4a). Supplementary Note 4 provides a more formal framework for the proposed in-memory parallel computation of CNF pseudo gradients.

Finally, another variation (Fig. S6) is due to the use of crossbar arrays based on active, three-terminal memory devices, such as 1T1R (Fig. S1d) or floating gate (Fig. S1e) memory. An additional "gate" signal in such crossbar arrays (Fig. S6a) can be used to condition the dot-product terms, which in turn allows for computing the make and break backward passes using a single crossbar in a single step (instead of two as shown in Fig. S4e) for high-throughput gain computation of CNF-form Boolean functions.

**Experimental results**

The key functionality of gain computation in CNF-form Boolean logic functions was experimentally validated by solving a high-order combinatorial optimization problem (Figs. 2, 3). Specifically, the studied optimization problem is custom-generated uniform random 3-SAT with $N$=14 variables and $M$=64 clauses. The problem parameters were chosen to maximize the use of hardware resources and problem hardness. The WalkSAT/SKC algorithm[14], a state-of-the-art local search heuristic, was implemented to solve a 3-SAT problem. Such an algorithm repeatedly flips variables using information on their break values to converge to the solution. An unsatisfied clause is first selected randomly out of all unsatisfied clauses determined in a forward pass. Break values for all variables within the selected clause are computed in parallel in the backward pass. Then, a specific variable is flipped according to the algorithm heuristics – see Methods section for more details on the 3-SAT problem and algorithm.

The experiments were performed on prototype board that features several 1T1R TaO$_x$ memristor crossbar arrays, back-end-of-the-line monolithically integrated with 180 nm CMOS circuits implementing driving, sensing, memory programming, and input/output (analog-to-digital conversion) functions. The forward and backward operations were tested on $M \times 2N$ sub-arrays of two crossbar arrays of a chip. Specifically, the conductance of on-state ($b_{ji}$=1) memristors were first tuned to $G_{on}$ = 130μS with 15% tuning accuracy using the write-verify approach (Fig. 2c), while the conductance of all off-state ($b_{ji}$=0) and outside of the utilized sub-array memristors were set to as small as possible (<10 μS) values. A single iteration of the algorithm involves the application of digital voltages $V_{xi} \equiv x_i V_0$ and $V'_{xi} \equiv \bar{x}_i V_0$, with $V_0$ = 0.2V, encoding literals $l_{2i-1} \equiv x_i$ and $l_{2i} \equiv \bar{x}_i$, correspondingly, to the word lines of the crossbar array for the forward computation (Figs. 2b and S5c). The bit-lines are tied to the ground so that the clause output currents $I_{cj}$ are computed in memory according to Kirchhoff's and Ohm's laws and correspond to dot-products $I_0 \sum_{i=1}^{2N} l_i b_{ji}$, where $I_0 \equiv V_0 G_{on}$ is a unit current via on-state coupling weight while assuming negligible off-state currents ($G_{off}$ =0), and $b_{ji}$ are coupling binary weights between literals $l_i$ and $j$-th clause. In the backward computation step, the word line voltages ($V_{cj}$) are applied to the second arrays' rows corresponding to the identified break clauses, while biasing other rows to zero. The bit line currents of the second array are multiplied by the corresponding literal values and normalized to the unit current $I_0$ to compute break values of variable $x_i$ (Fig. 2b and S5e), such that $\text{break}_{xi} = (x_i I_{xi} + \bar{x}_i I'_{xi})/I_0 = l_{2i-1} \sum_j b_{j,2i-1} z_j + l_{2i} \sum_j b_{j,2i} z_j$. Here $I_{xi}$ and $I'_{xi}$ denote the currents flowing in the bit lines corresponding to the normal ($l_{2i-1}$) and complimentary ($l_{2i}$) literals of variable $x_i$. Note that in the performed experiment, the bit line currents are always converted to the corresponding digital voltages using on-chip circuitry and transmitted out of the chip via a serial peripheral interface so that break clause checks and encode function (Fig. 2b) in the forward pass and heuristics for selecting a variable in the backward pass are performed on the personal computer (Fig. 2c).

The experimental demonstration was successful despite hardware nonidealities such as inaccurate tuning of on-state coupling weight conductances (Fig. 2c). While measured clause output currents (Fig. 3a) in the forward pass and break output currents (Fig. 3b) in the backward pass deviated from their ideal values, the margins between adjacent clause currents in the forward pass were sufficient to clearly distinguish break clauses – see, e.g., nonoverlapping histograms for 0, 1, and 2 clause current cases in Fig. 3a. The margins were slimmer for backward pass (Fig. 3b), though still large enough for correct operation, in part due to the stochastic nature of the convergence (Fig. 3c). Notably, the run-time-distribution, i.e. the cumulative probability of finding the solution over algorithm runtime[45] in hardware, follows closely the simulated one (Fig. 3d).

**Discussion**

The proposed approach is extremely compact, requiring ~$3NM$ total memory devices in the crossbar arrays, which can be used as a proxy for the overall hardware complexity, for massively parallel high-throughput computation of pseudo gradient in a polynomial with $M$ monomials and $N$ binary variables (Figs. 1 and S4e). Similarly, an efficient design for computing variable gains in CNF Boolean function with $M$ clauses and $N$ variable features ~$6NM$ memory devices (Fig. S5), which can be further reduced to ~$4NM$ complexity for three-terminal memory device implementations (Fig. S6) – see a specific example of 1T1R circuit in Fig. S8. Figure 4 compares the crossbar area ratio between QUBO-converted problems, required for implementation with quadratic HNNs/IMs, and the proposed approach on the different SAT benchmarks to quantify the area advantage in the context of SAT solvers. The advantage grows with the problem order exponentially, which can be analytically derived for SATs based on $K$-input XOR Boolean functions, and, e.g., ~ 3100 for the largest degree problem from SAT2020 competition benchmark (Figs. 4 and S9). Note that the direct comparison to recent work on high-order IMs is challenging because of, for example, the lack of hardware implementation details in Ref. 27, and hardware implementation specific to hard 3-SAT problems in Ref. 31. Nevertheless, a crude analysis of the latter approach, extended to $K$-order problems with $M$ clauses (by assuming an implementation with $M$ x $N$ array of unit cells, each hosting $K$-1 $N$:1 multiplexers) reveals inferior worst case ~$KN^2M$ complexity scaling.

Moreover, Fig. 4 area advantage estimates are rather conservative. We expect a comparable area of peripheral circuitry of crossbar arrays in both quadratic HNNs/IMs and the proposed implementation. On the other hand, QUBO-converted problems feature more variables, e.g., by ~4.5× more for hard 3-SAT problems, and this coefficient grows quickly with problem order[6]. Therefore, additional overheads are expected due to the mapping of coupling weights onto physical crossbar arrays, whose dimensions would be constrained by IR drops. Furthermore, HNNs/IMs implementations require multi-bit weights, while the proposed approach needs only binary weights, hence coming with lower programming circuitry overhead and/or enabling more compact crossbar array circuits based on conventional digital memory technologies.

Similar advantages are expected for speed and energy efficiency based on the proposed approach. Assuming negligible currents from off-state memory cells, the forward pass delay is independent of $K$. Similar to ratioed logic[46], worst-case logarithmic scaling with $K$ is expected for the backward pass latency, i.e., very weak dependence of the function degree. On the system level, in the advanced process implementations, energy is largely dominated by data movement in high-performance computing circuits[47], including in-memory computing circuits[48], so energy consumption is expected to increase roughly proportional to linear circuit dimensions. Furthermore, the more compact circuitry for gradient computation could enable fitting SAT solver completely on a chip, further improving efficiency by cutting energy and latency taxing inter-chip communication overheads. Additionally, solving optimization problems in a native high order form has led to faster convergence[25,27], partly due to additional spurious minima in the energy landscape of QUBO-converted problems[25]. Therefore, we expect hardware time-to-solution and energy-to-solution to be improved due to more efficient hardware and faster algorithmic convergence.

The proposed approach can be extended to computing gradients of functions with real-valued variables (Fig. S10). Let's first note that in the simplest case of multi-linear polynomial functions, a partial derivative with respect to a given variable equals to a sum of monomials that such a variable is a member of, divided by the value of that variable (Fig. S10a). Partial derivatives are computed in parallel to follow these steps. Like the binary variable function, the first step involves an $N \times M$ crossbar array with similarly configured binary memory weights to compute in-memory monomial products. Due to real-valued variables, the products are computed differently by applying logarithmically encoded variables and exponentiating the outputs (Fig. S10b). Appropriately weighted and specific (to the considered variable) monomial terms are summed up in the second array based on the multi-bit memory devices. Finally, the outputs from the second array are divided by variable values. Naturally, the division operation requires the variable to be nonzero, and may require a preprocessing step of shifting the variable ranges. Generalization beyond multi-linear function requires analog weights in the first array, and adequately adjusting weights in the second array (Fig. S10d,e). Figure S10c shows prospective in-memory implementation based on floating gate memories. Notably, exponentiation and division could be implemented directly in the second array by operating a floating gate transistor in the sub-threshold regime.

It is worth noting that a forward pass described in Fig. S5c is similar to earlier work on clause evaluation with content addressable memory implementations [49-51]. Indeed, the critical operation in the content addressable memory computation is in-memory vector-by-matrix multiplication between binary weights and inputs that produce binary "match" outputs. However, our approach takes advantage of all the outputs, not just binary match values. This feature and, more importantly, backward pass are key novelties of our approach, enabling for the first time massively parallel in-memory computing of make and break values of CNF-form Boolean functions and, more generally, pseudo gradient and gradient computation in functions of binary and real-valued variables respectively.

Finally, we believe the proposed approach has even more considerable potential for the algorithms that take advantage of massively parallel gradient computation. For example, local search algorithms based on all variable lookups, such as G2WSAT[45], which is more powerful than WalkSAT/SKC, have been suggested; however, their applications seem to be limited due to high computational costs when running on conventional hardware. The efficient hardware for solving problems natively in the high-order space could also boost new efficient problem embeddings, such as for quadratic assignment problems, by eliminating commonly used one-hot encoding in favor of much more compact high-order problem formulations[39].

**Methods**

**3-SAT instance generation**

Random uniform 3-SAT problems with $M$ clauses and $N$ variables are generated by initializing an empty list of clauses and repeatedly adding "valid" clauses to that list, one at a time, until the number of clauses reaches the target value of $M$. Specifically, a "candidate" clause is generated by randomly sampling three out of $2N$ literals without replacement. Each of the $2N$ literals has an equal probability of being chosen. Two criteria are then checked to determine if the candidate clause is valid: If both literals of a variable are present in the candidate clause; and if the candidate clause is already present in the clause list. If both criteria are false, such a clause is added to the list. Once the entire list of clauses is ready, a SAT solver is run on the generated instance to check if it is satisfiable.

3-SAT problems with $M = 64$ clauses are considered. This number matches the linear dimension of the memristor crossbar arrays and hence corresponds to the largest SAT problems in terms of the number of clauses that can be implemented with an experimental setup without employing time-consuming time-multiplexing techniques. Prior work shows[13] that satisfiable randomly-generated uniform 3-SAT problems with 14 variables, corresponding to clause-to-variable ratio $M/N \approx 4.57$, are among the hardest. This was confirmed by generating and solving multiple instances of 3-SAT problems with $M = 64$ and $N$ in the range from 12 to 21.

Fig. S11 provides a specific instance that was solved in the experimental demonstration in the common "CNF" format.

**WalkSAT/SKC algorithm**

The implemented algorithm has the following structure:

**Input:** 3-SAT CNF-formula, MAX_FLIPS=10,000, MAX_ITER=1, probability $p$ =0.5

**Output:** "true", if a satisfying assignment is found, "false" otherwise

1. **for** $t$ = 1 to MAX_ITER
2.    Randomly initialize variable assignments $\mathbf{X} = \mathbf{X_0}$
3.    **for** $f$ = 1 to MAX_FLIPS
4.       **if** $\mathbf{X}$ is a solution, **return** true
5.       Randomly choose an unsatisfied clause $c$
6.       Calculate set of break values (**BV**) for all member variables of $c$
7.       **if** min(**BV**) = 0, flip variable with zero break value; pick randomly in a tiebreaker
8.       **else**
9.          with probability $p$ select & flip a variable in the clause randomly
10.          with probability 1-$p$ select & flip a variable with the smallest **break value**; pick at random in a tiebreaker

Note that similarly to the original version[14], known as WalkSAT/SKC, this algorithm uses only break values as deciding metric.

**Experimental setup**

The experimental setup consists of a custom chip hosting three 64×64 memristive crossbar arrays (Fig. S6c) integrated with the custom printed circuit board (PCB) (Fig. S6d), and custom-written firmware and Python scripts to communicate with the chip using software functions.

The Ta/TaO$_x$/Pt memristors were monolithically integrated in-house on CMOS circuits fabricated in a TSMC's 180 nm technology node (Fig. S6a,b). The integration starts with the removal of silicon nitride and oxide passivation from the surface of the CMOS wafer with reactive ion etching, and a buffered oxide etch dip. Chromium and platinum bottom electrodes are then patterned with e-beam lithography and metal lift-off process, followed by reactive sputtered 4.5 nm tantalum oxide as the switching layer. The device stack is finalized by e-beam lithography patterning of sputtered tantalum and platinum metal as top electrodes.

The chip's CMOS subsystem implements digital control and analog sensing circuits for performing in-memory analog computations (Fig. S6e). Each array utilizes digital-to-analog converters (DACs) to drive analog voltages (inputs) to the rows (i.e., word and gate lines) of the array. There are transimpedance amplifiers (TIAs) followed by sample-and-hold (S&H) circuits at the outputs to rapidly convert the currents to voltages, and sample them while providing virtual ground to the column (bit) lines. The sampled voltages are then multiplexed and converted to digital values using analog-to-digital converters (ADCs), each shared by 16 columns.

The PCB supplies DC analog reference signals to the chip, hosts a microcontroller, and provides a digital interface between the chip and the Python scripts running on a personal computer via serial communication. Ref. 52 provides more information on the memristor fabrication and its integration with CMOS circuits.

**Acknowledgments**

This material is based upon work supported by the Defense Advanced Research Projects Agency (DARPA) under the Air Force Research Laboratory (AFRL) Agreement No. FA8650-23-3-7313. The views, opinions, and/or findings expressed are those of the authors and should not be interpreted as representing the official views or policies of the Department of Defense or the U.S. Government.

# Figures

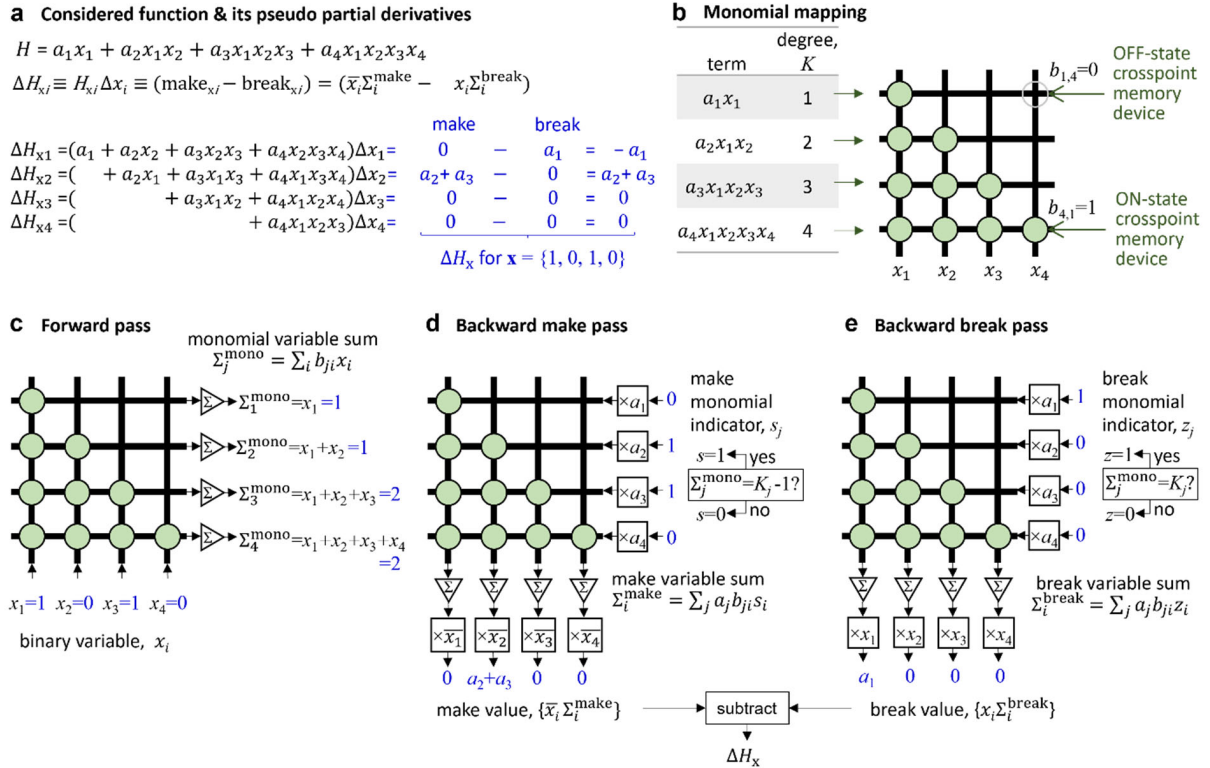

**Fig. 1. Pseudo gradient computation for binary-variable functions. a.** Considered 4$^{th}$-degree polynomial function. Note that binary-variable high-order polynomials are multilinear, i.e., degree of any variable in a term is not more than one. **b-e.** Main idea of the proposed approach showing **(b)** crossbar memory array implementation and **(c-e)** three in-memory computing operations for parallel gradient computation. **c.** Sums of monomial variables are first computed in the forward vector-by-matrix multiplication pass. These values are compared to monomial orders in backward steps **(d)** and **(e)** to identify break and make type monomials. Then, unit inputs, scaled by a monomial factor, are applied for the identified break and make monomials in the backward vector-by-matrix multiplication. Finally, the results are appropriately gated at the periphery to compute make and break terms of the pseudo partial derivatives for each variable. The partial derivatives in question correspond to a difference between make and break components, as shown in panels

a, d, and e. Values shown in blue correspond to specific variable assignment $x_1=1$, $x_2=0$, $x_3=1$, $x_4=0$.

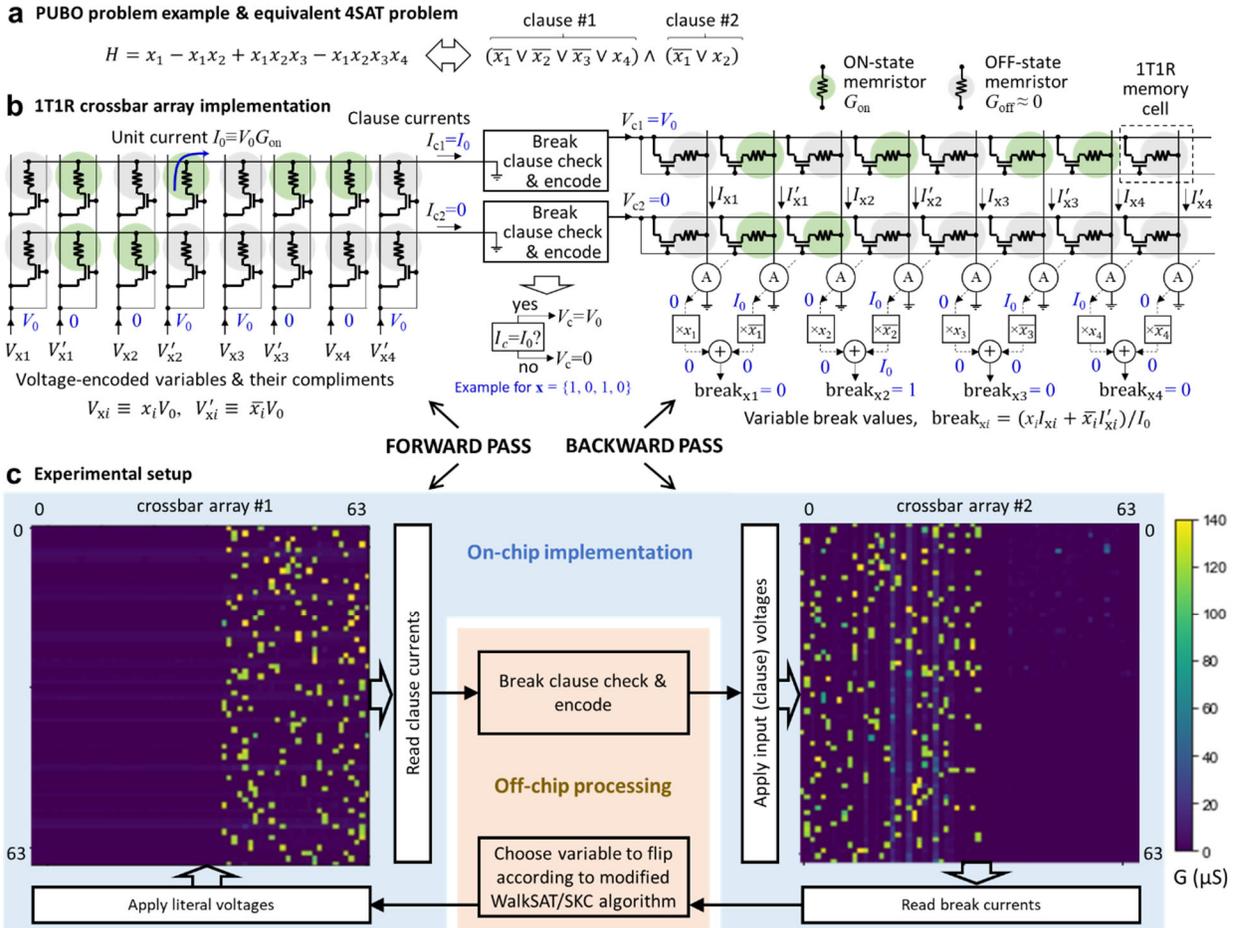

**Fig. 2. In-memory computing hardware implementation. a.** A toy example of CNF Boolean function with $M=2$, $N=K=4$, which is equivalent to the monomial in Fig. 1a with $a_1 = 1$, $a_2 = -1$, $a_3 = 1$, and $a_4 = -1$ when represented in a PUBO form. **b.** A prospective in-memory 1T1R memristor circuit implementation of panel a problem for parallel computation of its break values. Each 1T1R memory cell is comprised of a select transistor coupled with memristor. Note the gate voltages are tied to the word lines in the utilized chip to suppress leakages through unselected memory cells. A text shown in blue corresponds to a specific variable assignment $x_1=1$, $x_2=0$, $x_3=1$, $x_4=0$. **c.** Experimental setup details for solving the considered 14-variable 64-clause 3-SAT problem. Two arrays represent 1T1R crossbar circuits of a chip in the integrated CMOS/memristor setup and are used for demonstrating forward and backward passes. The colormaps show the measured conductance of programmed memristors corresponding to the studied 3-SAT problem in the experiment.

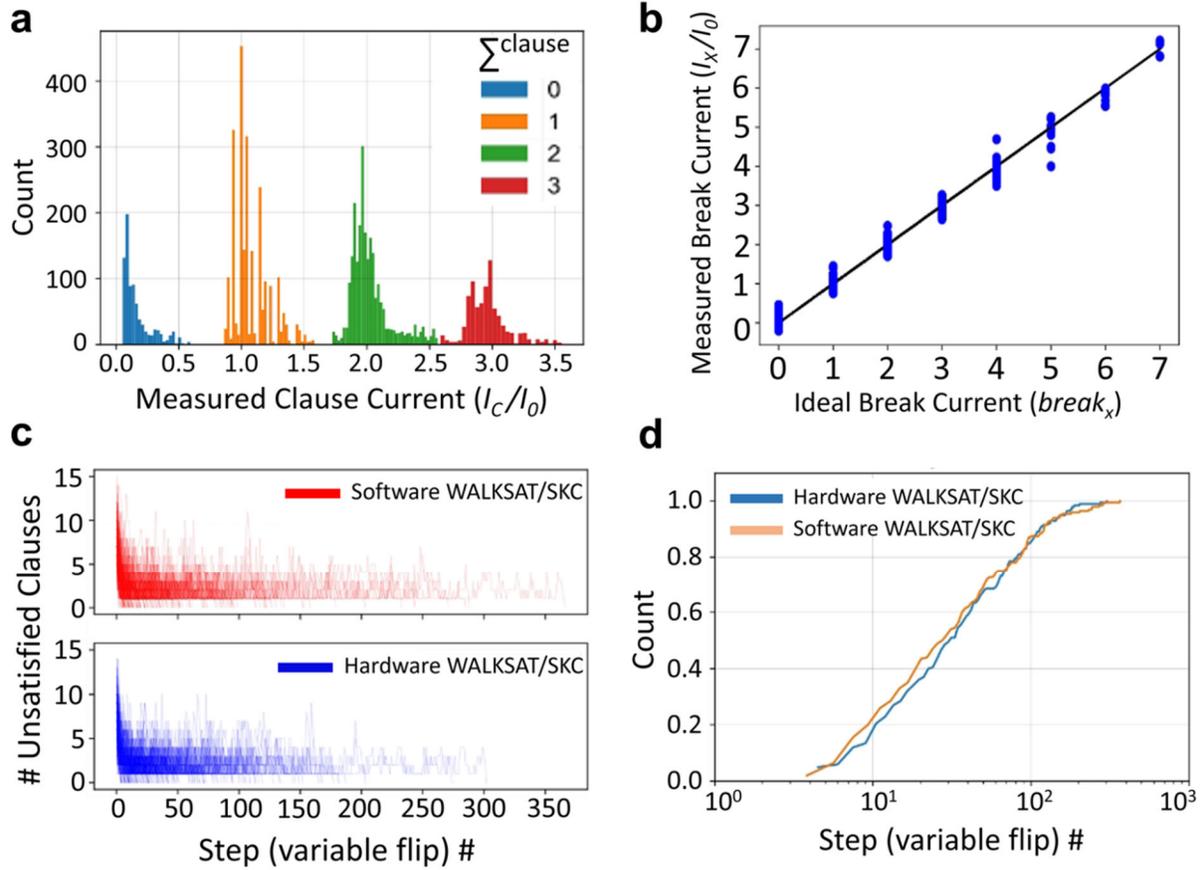

**Fig. 3. Experimental results**. **a,b** Measured output (bit-line) currents for (**a**) forward and (**b**) backward passes. In panel **a**, currents are grouped according to their ideal clause literal sum values. In panel **b**, outputs are shown against calculated ideal break values. In both panels, data are collected during a single iteration of an algorithm. **c, e** Functional performance comparison between ideal software model and experiment, showing (**c**) evolution of the number of unsatisfied clauses, and (**d**) run-length distribution curves obtained across 200 iterations (restarts) of the algorithm, each time with new randomly initialized variable assignments. In all experiments, MAX_FLIPS = 10,000.

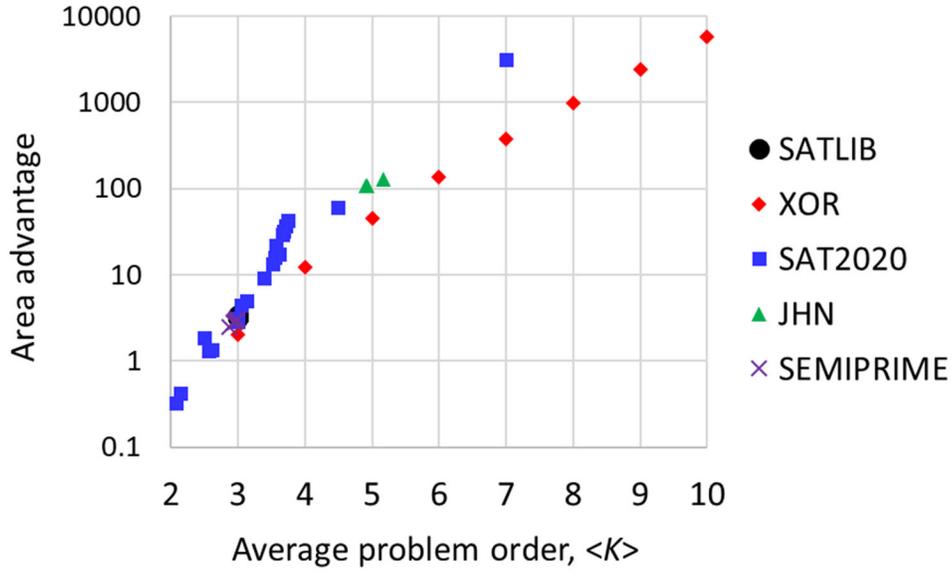

**Fig. 4. SAT solver area advantage for the proposed approach over quadratic HNN/IM.** The area advantage is defined as the ratio of the total number of memory devices in the crossbar arrays required for both implementations. The studied benchmark problems SATLIB, XOR, SAT2020, JHN, and SEMIPRIME correspond to, respectively, uniform random 3-SAT problems[53], custom-generated $K$-input XOR problems, competition 3-SAT problems[54], "JHN" DIMACS benchmark instances from SATLIB benchmark[53], and custom-generated 3-SAT problems for semiprime factoring[55]. The higher order problems were first converted to 3-SAT with order-reduction technique[31] and then converted to corresponding QUBO problems using the Rosenberg approach[32]. Note that a more compact QUBO formulation can be obtained for XOR and other problems by using Tseytin transformation[56], though at the cost of substantial preprocessing overhead. Fig. S9 shows the data used for this figure.

# Computing High-Degree Polynomial Gradients in Memory

T. Bhattacharya[1*], G. Hutchinson[1], G. Pedretti[2], X. Sheng,[2] J. Ignowski[2], T. Van Vaerenbergh[3], R. Beausoleil[3], J.P. Strachan[4] & D.B. Strukov[1*]

## Supplementary Information

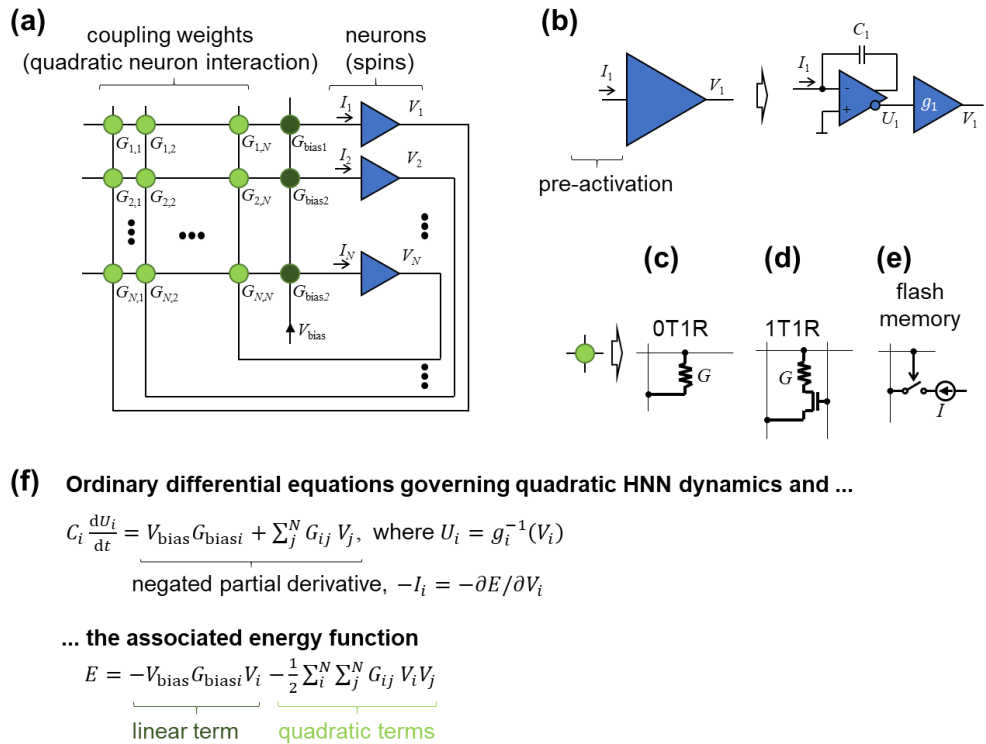

**Figure S1. Quadratic Hopfield neural network:** (a) Neural network structure, (b) examples of neuron and (c-e) synapse implementations with (c,d) memristive and (e) flash memory device technologies, and (f) equations describing network operation. A shown neuron implementation is based on a differential-output operational amplifier (opamp) with capacitive feedback, whose noninverting output is connected to the second opamp implementing nonlinear activation, e.g., sigmoid function. Due to virtual ground configuration, the current to the first opamp is effectively neuron pre-activation, i.e., $I_i = \sum G_{ij} V_j + G_{biasi} V_{bias}$. In panel c, "0T1R" stands for zero transistors plus one resistive switching element (memristor) per memory cell, while in panel d, "1T1R" stands for one transistor plus one memristor. Note that the figure shows HNN with type I dynamics according to classification in Ref. 29, i.e., does not include $-U_i$ "leakage" term in the right-hand side of differential equations that are common to a more widespread HNN implementation with dynamics type II. Also, in principle, type I dynamics implies unbounded values of $U$, while they would be bounded in the practical hardware implementation. This introduces a nonlinear "squishing" function for $U$ that is, for simplicity, not shown in panel f equations. However, such an additional squishing function does not modify the minima of the original energy function.

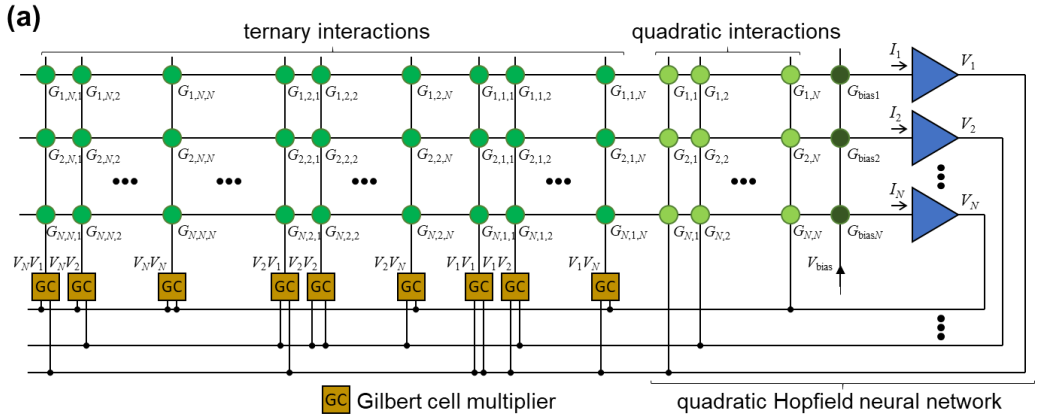

**Figure S2. High-order Hopfield neural network**: (a) The network structure and (b) equations governing operation. For clarity, the figure shows a third-order HNN (of type I, similar to Figure S1) with a regular implementation at the cost of some redundancy, which is chosen to highlight how such third-order ($K=3$) HNN can be extended to higher-order ($K>3$) networks. Specifically, the shown implementation of third-order interactions employs $N^2$ columns and hence $N^3$ synapses. Because of the symmetric coupling matrix and typically employed multi-linear energy functions, the number of columns can be reduced to $(N-1)N/2$, while the number of synapses to $(N-1)N^2/2$ in a more optimal implementation for third-order HNNs.

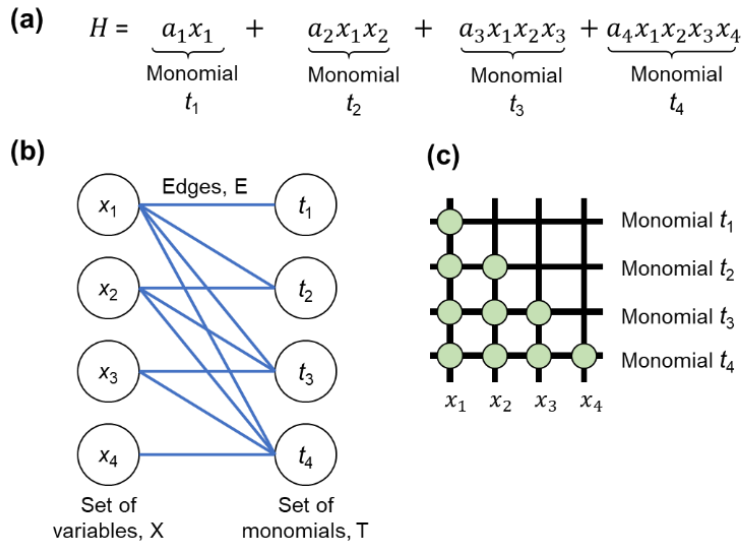

**Figure S3. Mapping polynomial to the crossbar array:** (a) Considered multi-linear polynomial example, (b) its graph representation, and (c) corresponding crossbar circuit implementation.

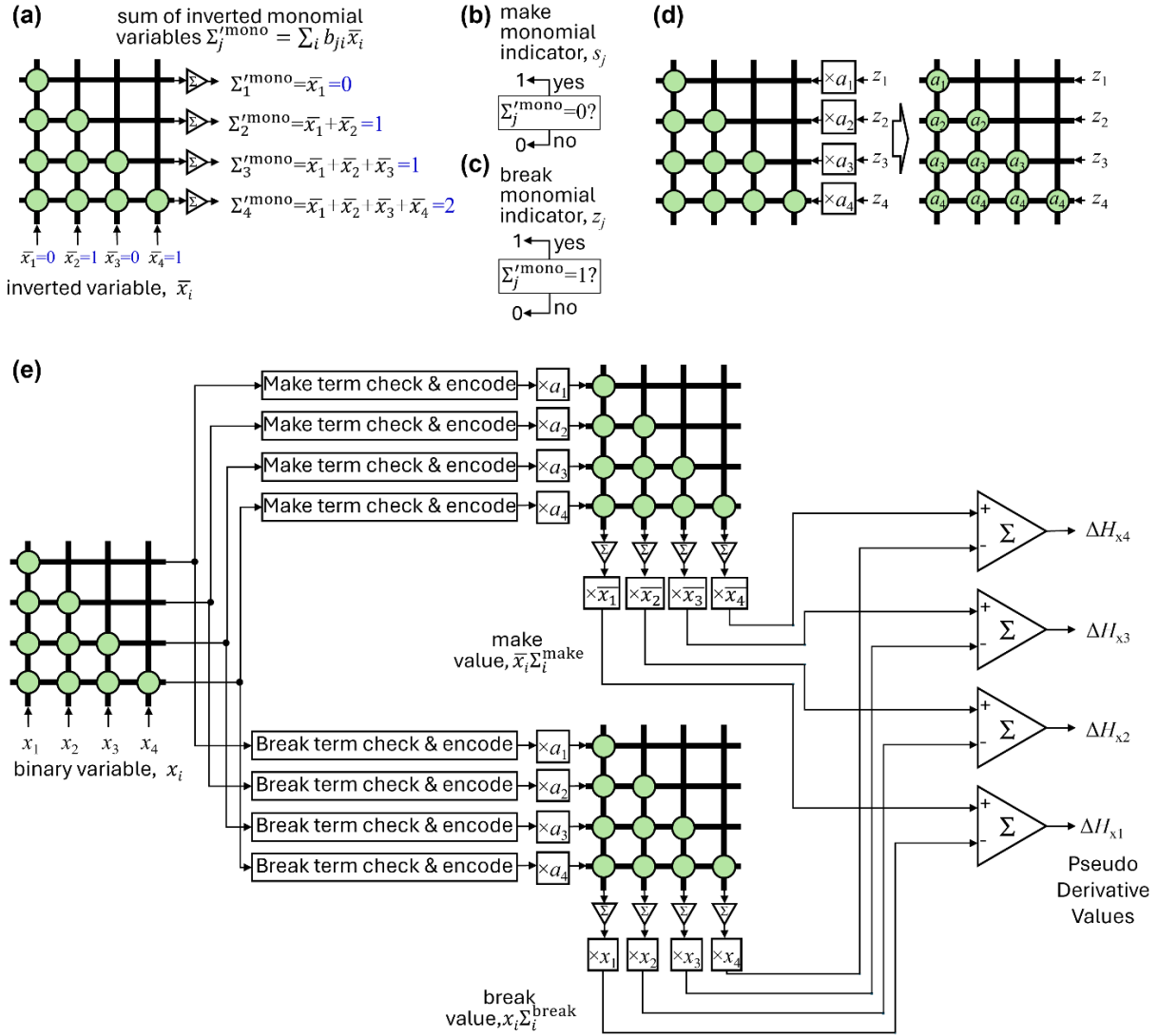

**Figure S4. Simplified periphery and high-throughput extensions:** (a-d) Simplifying peripheral circuitry by (a) inverting variable inputs in the forward pass and (d) implementing constant multiplication with multi-valued memory devices in the backward passes. With such modification, identifying make and break monomials with inverted variable inputs requires comparing dot products of the forward pass with fixed values of (b) 0 and (c) 1, correspondingly. (e) Increasing throughput by implementing backward passes with separate crossbar memory arrays. On panel a, values in blue correspond to the inverted version of the same variable assignment $x_1=1$, $x_2=0$, $x_3=1$, $x_4=0$ as that was used in Fig. 1a of main text. The polynomial mapped to the crossbars in all panels is the same as that in Fig. 1a of main text.

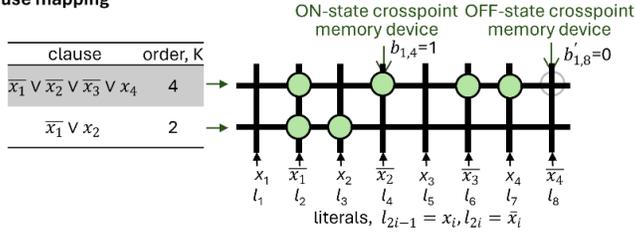
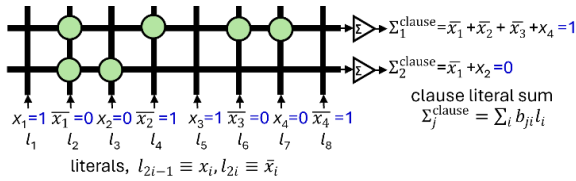
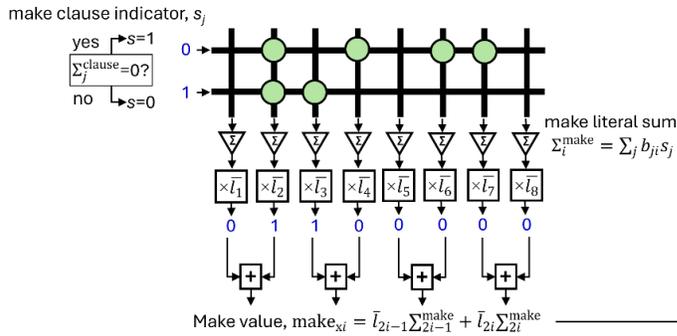
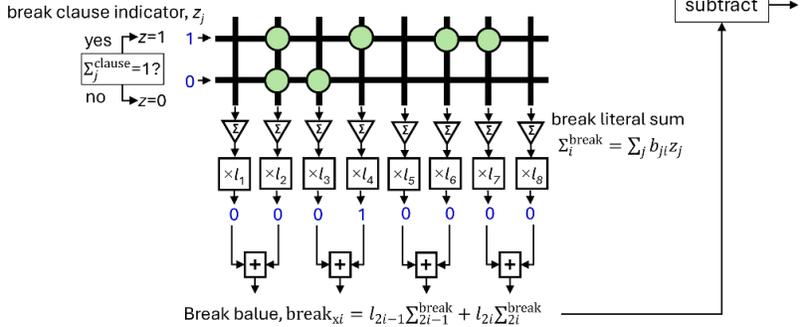

**Figure S5. Extension for computing CNF gains:** (a) Considered 4th-degree CNF Boolean function, i.e., 4-SAT problem. (b-e) The main idea of the gain computing showing (b) crossbar memory array implementation and (c-e) three in-memory computing operations. The approach is similar to the polynomial pseudo gradient computation. Specifically, sums of clause literals are first computed similarly to the inverted monomial approach (Fig. S4a) in the forward vector-by-matrix multiplication pass as shown in panel (c). The sum values are compared to clause orders in backward steps (d) and (e) to identify break and make type clauses. Unit inputs, that can be scaled by clause factor when clauses in CNF are weighted (e.g., used in weighted SAT problems), are applied for the identified break and make clauses in the backward vector-by-matrix multiplication. The results are then properly gated at the periphery to compute make and break values for each variable. The corresponding variable gains (i.e., negative pseudo partial derivatives) are found by subtracting the outputs of two backward passes as shown in panel a. Values shown in blues correspond to $x_1=1$, $x_2=0$, $x_3=1$, $x_4=0$, i.e., the same example of assignment as in Fig. 1. Also note that the considered 4-SAT problem is equivalent to polynomial function in Fig. 1a, when assuming $a_1 = 1$, $a_2 = -1$, $a_3 = 1$, and $a_4 = -1$.

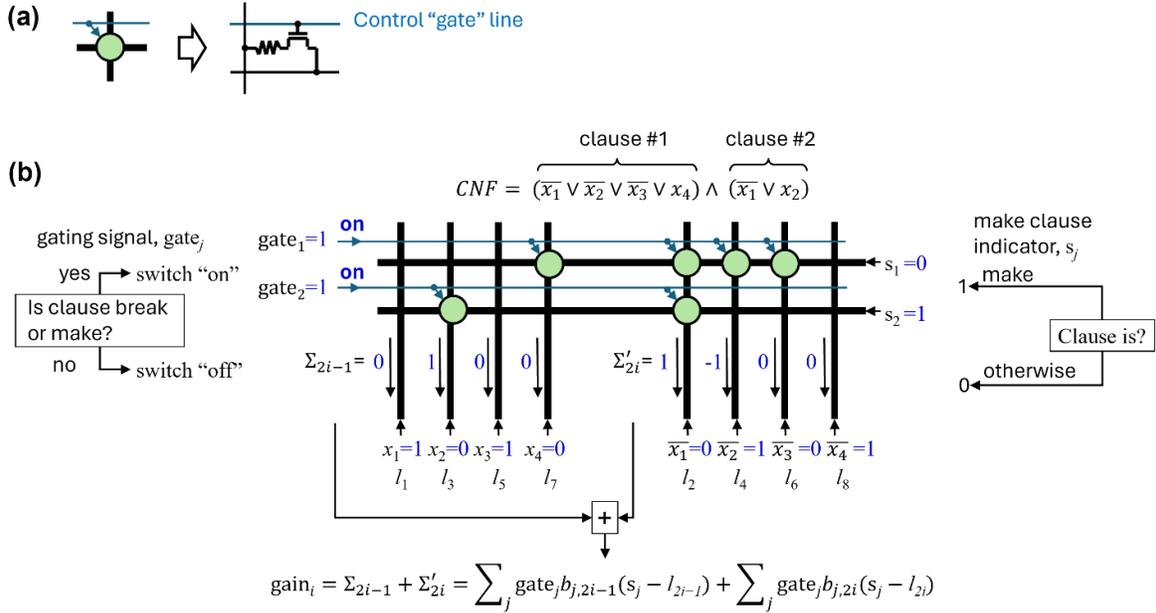

**Figure S6. Extension for three-terminal memory devices:** (a) A three-terminal memory device based on 1T1R memory. (b) Backward pass implementation to compute gain values using a single $M \times 2N$ crossbar array (instead of two such arrays as shown in Figs. S5d and S5e), where $M$ is the number of clauses and $N$ is the number of variables. Values in blue correspond to the specific variable assignment $x_1=1$, $x_2=0$, $x_3=1$, $x_4=0$. $\Sigma_{2i-1}$ and $\Sigma'_{2i}$ are sums of triple products along $(2i\text{-}1)$-th and $2i$-th columns corresponding to variable $x_i$'s normal and complementary literals. Specifically, the triple product in $\Sigma_{2i-1}$ ( $\Sigma'_{2i}$ ) consists of terms $s_j - l_{2i-1}$ ($s_j - l_{2i}$), where $s_j$ is the make clause indicator, the weight value $b_{j,2i-1}$ ($b_{j,2i}$), and gating signal $gate_j$. The gating signals are applied to 1T1R control lines and turn on only make or break type clauses. Variable gain values are obtained by adding $\Sigma_{2i-1}$ and $\Sigma'_{2i}$ - see Eq. 47 of Supplementary Note 4 for more details.

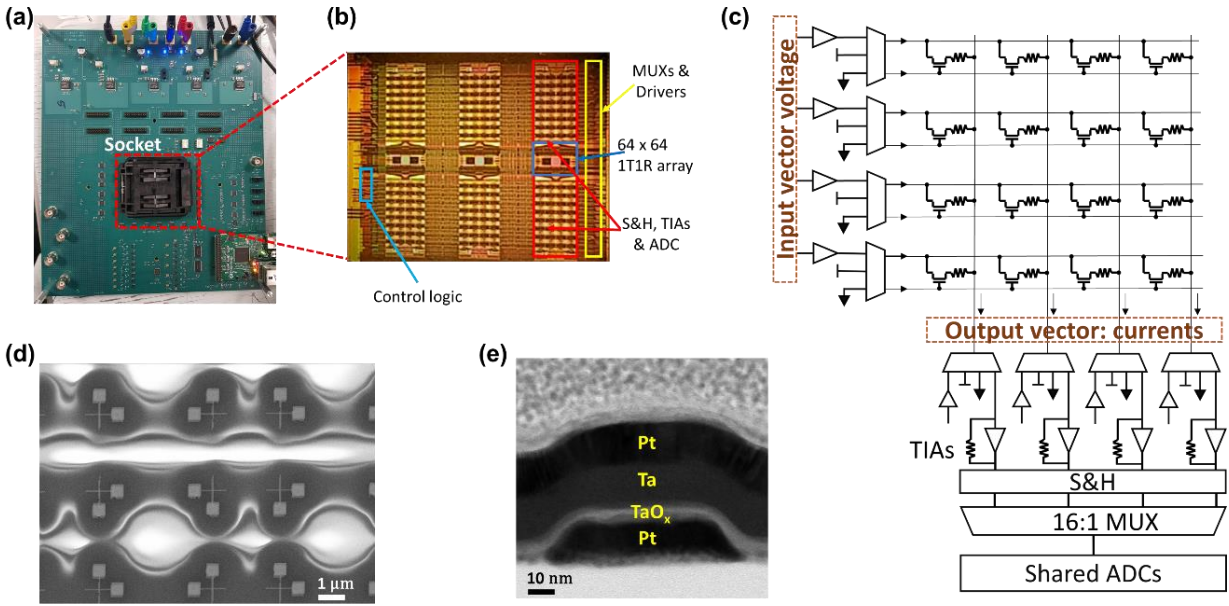

**Figure S7. Experimental setup.** Hewlett Packard Enterprise's CMOS/memristor prototyping setup[52]. (a) Photo of the printed circuit board used for testing the chip. (b) Photo of a CMOS die hosting three 1T1R memristor crossbar arrays and peripheral circuitry. (c) Circuit schematic of a single crossbar and its peripherals from the CMOS/memristor chip. (d) Scanning Electron Microscope (SEM) image of the memristor devices and (e) cross-section Transmission Electron Microscopy (TEM) image of the Pt/Ta/TaOx/Pt memristor stack.

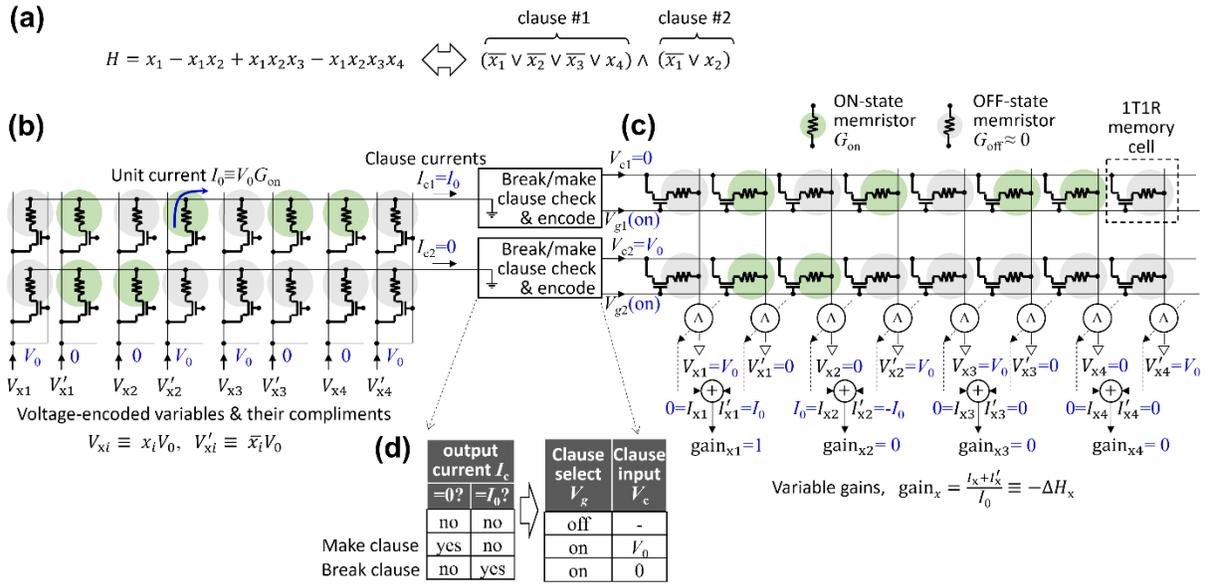

**Figure S8. A compact 1T1R implementation for CNF-form Boolean function gain calculation.** (a) The considered toy example of CNF function. (b, c) 1T1R circuit implementations of (b) forward and (c) backward computation of the proposed approach. (d) Peripheral logic functionality. The bit lines in the second array are tied to the literal values so that $I_x + I'_x$ is proportional to the $\text{gain}_{xi} = \sum_j \text{gate}_j b_{j,2i-1}(s_j - l_{2i-1}) + \sum_j \text{gate}_j b_{j,2i}(s_j - l_{2i}) -$ see Fig. S6, its discussion, and Supplementary Note 4. Panels a and b are the same as in Fig. 2a and are shown for convenience. Also, note that backward pass in such an approach requires applying different voltages to the word, control and bit lines of the array, which cannot be implemented in the utilized experimental setup.

| | Benchmark problem | # var (N) | # clauses (M) | Clause /Var (M/N) | Maximum order Kmax | Average order <K> | # var after converting to QUBO | # weights in proposed approach | # weights in QUBO with Rosenberg mapping | Ratio |
|---|---|---|---|---|---|---|---|---|---|---|
| SATLIB | uf20-01.cnf | 20 | 91 | 4.55 | 3 | 3.00 | 111 | 7.28E+03 | 2.46E+04 | 3.38 |
| | uf50-01.cnf | 50 | 218 | 4.36 | 3 | 3.00 | 268 | 4.36E+04 | 1.44E+05 | 3.29 |
| | uf75-01.cnf | 75 | 325 | 4.33 | 3 | 3.00 | 400 | 9.75E+04 | 3.20E+05 | 3.28 |
| | uf100-01.cnf | 100 | 430 | 4.30 | 3 | 3.00 | 530 | 1.72E+05 | 5.62E+05 | 3.27 |
| | uf125-01.cnf | 125 | 538 | 4.30 | 3 | 3.00 | 663 | 2.69E+05 | 8.79E+05 | 3.27 |
| | uf150-01.cnf | 150 | 645 | 4.30 | 3 | 3.00 | 795 | 3.87E+05 | 1.26E+06 | 3.27 |
| | uf175-01.cnf | 175 | 753 | 4.30 | 3 | 3.00 | 928 | 5.27E+05 | 1.72E+06 | 3.27 |
| | uf200-01.cnf | 200 | 860 | 4.30 | 3 | 3.00 | 1060 | 6.88E+05 | 2.25E+06 | 3.27 |
| | uf225-01.cnf | 225 | 960 | 4.27 | 3 | 3.00 | 1185 | 8.64E+05 | 2.81E+06 | 3.25 |
| | uf250-01.cnf | 250 | 1065 | 4.26 | 3 | 3.00 | 1315 | 1.07E+06 | 3.46E+06 | 3.25 |
| XOR | xor3.cnf | 3 | 4 | 1.33 | 3 | 3.00 | 7 | 4.80E+01 | 9.80E+01 | 2.04 |
| | xor4.cnf | 4 | 8 | 2.00 | 4 | 4.00 | 28 | 1.28E+02 | 1.57E+03 | 12.25 |
| | xor5.cnf | 5 | 16 | 3.20 | 5 | 5.00 | 85 | 3.20E+02 | 1.45E+04 | 45.16 |
| | xor6.cnf | 6 | 32 | 5.33 | 6 | 6.00 | 230 | 7.68E+02 | 1.06E+05 | 137.76 |
| | xor7.cnf | 7 | 64 | 9.14 | 7 | 7.00 | 583 | 1.79E+03 | 6.80E+05 | 379.34 |
| | xor8.cnf | 8 | 128 | 16.00 | 8 | 8.00 | 1416 | 4.10E+03 | 4.01E+06 | 979.03 |
| | xor9.cnf | 9 | 256 | 28.44 | 9 | 9.00 | 3337 | 9.22E+03 | 2.23E+07 | 2416.57 |
| | xor10.cnf | 10 | 512 | 51.20 | 10 | 10.00 | 7690 | 2.05E+04 | 1.18E+08 | 5775.01 |
| SAT2020 | battleship-3-5-sat.cnf | 15 | 24 | 1.60 | 5 | 3.13 | 60 | 1.44E+03 | 7.20E+03 | 5.00 |
| | battleship-4-7-sat.cnf | 28 | 58 | 2.07 | 7 | 3.38 | 172 | 6.50E+03 | 5.92E+04 | 9.11 |
| | battleship-5-9-sat.cnf | 45 | 115 | 2.56 | 9 | 3.52 | 370 | 2.07E+04 | 2.74E+05 | 13.23 |
| | battleship-6-10-sat.cnf | 60 | 186 | 3.10 | 10 | 3.55 | 600 | 4.46E+04 | 7.20E+05 | 16.13 |
| | battleship-6-11-sat.cnf | 66 | 201 | 3.05 | 11 | 3.61 | 678 | 5.31E+04 | 9.19E+05 | 17.33 |
| | Steiner-9-4-bce.cnf | 63 | 55 | 0.87 | 3 | 2.56 | 95 | 1.39E+04 | 1.81E+04 | 1.30 |
| | Steiner-15-6-bce.cnf | 165 | 146 | 0.88 | 3 | 2.60 | 254 | 9.64E+04 | 1.29E+05 | 1.34 |
| | sgen1-sat-100-100.cnf | 100 | 240 | 2.40 | 5 | 2.50 | 300 | 9.60E+04 | 1.80E+05 | 1.88 |
| | sgen3-n120-s12930489-sat.cnf | 120 | 288 | 2.40 | 5 | 2.50 | 360 | 1.38E+05 | 2.59E+05 | 1.88 |
| | sgen1-sat-120-100.cnf | 120 | 288 | 2.40 | 5 | 2.50 | 360 | 1.38E+05 | 2.59E+05 | 1.88 |
| | sgen1-sat-140-100.cnf | 140 | 336 | 2.40 | 5 | 2.50 | 420 | 1.88E+05 | 3.53E+05 | 1.88 |
| | sgen4-sat-160-8.cnf | 160 | 384 | 2.40 | 5 | 2.50 | 480 | 2.46E+05 | 4.61E+05 | 1.88 |
| | sgen1-sat-160-100.cnf | 160 | 384 | 2.40 | 5 | 2.50 | 480 | 2.46E+05 | 4.61E+05 | 1.88 |
| | driverlog1_ks99i.renamed-as.sat05-3951.cnf | 207 | 588 | 2.84 | 4 | 2.08 | 281 | 4.87E+05 | 1.58E+05 | 0.32 |
| | mod2-rand3bip-sat-210-2.sat05-2159.reshuffled-07.cnf | 210 | 840 | 4.00 | 3 | 3.00 | 1050 | 7.06E+05 | 2.21E+06 | 3.13 |
| | mod2-rand3bip-sat-220-2.sat05-2174.reshuffled-07.cnf | 220 | 880 | 4.00 | 3 | 3.00 | 1100 | 7.74E+05 | 2.42E+06 | 3.13 |
| | mod2-rand3bip-sat-240-3.sat05-2205.reshuffled-07.cnf | 240 | 960 | 4.00 | 3 | 3.00 | 1200 | 9.22E+05 | 2.88E+06 | 3.13 |
| | driverlog3_v01a.renamed-as.sat05-3963.cnf | 170 | 1559 | 9.17 | 5 | 2.14 | 472 | 1.06E+06 | 4.46E+05 | 0.42 |
| | mod2c-rand3bip-sat-190-3.sat05-2445.reshuffled-07.cnf | 271 | 1972 | 7.28 | 6 | 4.48 | 8091 | 2.14E+06 | 1.31E+08 | 61.25 |
| | 20180322_164245263_p_cnf_320_1120.cnf | 320 | 1120 | 3.50 | 3 | 3.00 | 1440 | 1.43E+06 | 4.15E+06 | 2.89 |
| | 289-sat-11x4.cnf | 176 | 1628 | 9.25 | 4 | 3.68 | 4268 | 1.15E+06 | 3.64E+07 | 31.79 |
| | 289-sat-4x8.cnf | 128 | 896 | 7.00 | 4 | 3.57 | 2240 | 4.59E+05 | 1.00E+07 | 21.88 |
| | 289-sat-5x8.cnf | 160 | 1400 | 8.75 | 4 | 3.66 | 3640 | 8.96E+05 | 2.65E+07 | 29.58 |
| | 289-sat-6x8.cnf | 192 | 2016 | 10.50 | 4 | 3.71 | 5376 | 1.55E+06 | 5.78E+07 | 37.33 |
| | 289-sat-6x9.cnf | 216 | 2538 | 11.75 | 4 | 3.74 | 6858 | 2.19E+06 | 9.41E+07 | 42.90 |
| | 289-sat-7x6.cnf | 168 | 1554 | 9.25 | 4 | 3.68 | 4074 | 1.04E+06 | 3.32E+07 | 31.79 |
| | **7cnf20_90000_90000_7.shuffled.cnf** | **20** | **1532** | **76.60** | **7** | **7.00** | **13808** | **1.23E+05** | **3.81E+08** | **3111.31** |
| | aes_32_1_keyfind_1.cnf | 300 | 1016 | 3.39 | 5 | 3.05 | 1656 | 1.22E+06 | 5.48E+06 | 4.50 |
| | aes_32_1_keyfind_2.cnf | 300 | 1016 | 3.39 | 5 | 3.05 | 1656 | 1.22E+06 | 5.48E+06 | 4.50 |
| JHN | jnh1.cnf | 100 | 850 | 8.50 | 14 | 5.17 | 4697 | 3.40E+05 | 4.41E+07 | 129.78 |
| | jnh2.cnf | 100 | 850 | 8.50 | 10 | 4.93 | 4311 | 3.40E+05 | 3.72E+07 | 109.32 |
| | jnh3.cnf | 100 | 850 | 8.50 | 11 | 4.90 | 4265 | 3.40E+05 | 3.64E+07 | 107.00 |
| SEMIPRIME | semiprime15.cnf | 22 | 73 | 3.32 | 3 | 2.86 | 90 | 6.42E+03 | 1.62E+04 | 2.52 |
| | semiprime21.cnf | 52 | 187 | 3.60 | 3 | 2.93 | 232 | 3.89E+04 | 1.08E+05 | 2.77 |
| | semiprime55.cnf | 68 | 248 | 3.65 | 3 | 2.94 | 308 | 6.75E+04 | 1.90E+05 | 2.81 |
| | semiprime91.cnf | 116 | 434 | 3.74 | 3 | 2.95 | 540 | 2.01E+05 | 5.83E+05 | 2.90 |
| | semiprime221.cnf | 138 | 519 | 3.76 | 3 | 2.96 | 646 | 2.86E+05 | 8.35E+05 | 2.91 |
| | semiprime323.cnf | 204 | 777 | 3.81 | 3 | 2.97 | 968 | 6.34E+05 | 1.87E+06 | 2.96 |
| | semiprime651.cnf | 232 | 886 | 3.82 | 3 | 2.97 | 1104 | 8.22E+05 | 2.44E+06 | 2.96 |
| | semiprime1271.cnf | 316 | 1216 | 3.85 | 3 | 2.97 | 1516 | 1.54E+06 | 4.60E+06 | 2.99 |
| | semiprime8633.cnf | 492 | 1908 | 3.88 | 3 | 2.98 | 2380 | 3.75E+06 | 1.13E+07 | 3.02 |
| | semiprime60491.cnf | 658 | 2563 | 3.90 | 3 | 2.98 | 3198 | 6.75E+06 | 2.05E+07 | 3.03 |

**Figure S9. Area advantage for the studied benchmarks.** The highlighted row corresponds to the largest area advantage for the SAT2020 industrial benchmark problem.

**(a)**

$H = a_1x_1 + a_2x_1x_2 + a_3x_1x_2x_3 + a_4x_1x_2x_3x_4$

$H_{x1} = (a_1x_1 + a_2x_1x_2 + a_3x_1x_2x_3 + a_4x_1x_2x_3x_4)/x_1$
$H_{x2} = (\quad\quad a_2x_1x_2 + a_3x_1x_2x_3 + a_4x_1x_2x_3x_4)/x_2$
$H_{x3} = (\quad\quad\quad\quad\quad\quad a_3x_1x_2x_3 + a_4x_1x_2x_3x_4)/x_3$
$H_{x4} = (\quad\quad\quad\quad\quad\quad\quad\quad\quad\quad a_4x_1x_2x_3x_4)/x_4$

**(b)** 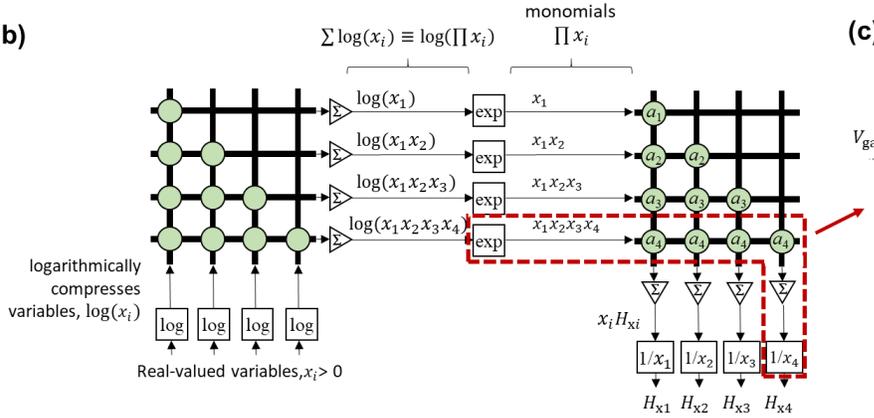

**(c)** 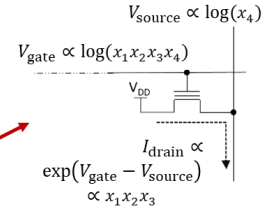

**(d)**

$H = a_1x_1 + a_2x_1x_2 + a_3x_1x_2x_3 + a_4x_1x_2x_3x_4^3$

$H_{x1} = (a_1x_1 + a_2x_1x_2 + a_3x_1x_2x_3 + a_4x_1x_2x_3x_4^3)/x_1$
$H_{x2} = (\quad\quad a_2x_1x_2 + a_3x_1x_2x_3 + a_4x_1x_2x_3x_4^3)/x_2$
$H_{x3} = (\quad\quad\quad\quad\quad\quad a_3x_1x_2x_3 + a_4x_1x_2x_3x_4^3)/x_3$
$H_{x4} = (\quad\quad\quad\quad\quad\quad\quad\quad\quad\quad 3a_4x_1x_2x_3x_4^3)/x_4$

**(e)** 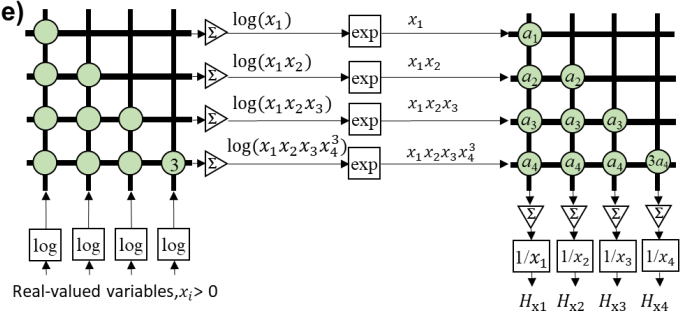

**Figure S10. Gradient computation of real-valued functions:** (a) Considered multi-linear polynomial energy function of real-valued variables and its partial derivatives. (b) The main idea and (c) circuit implementation with floating gate memories. In such an implementation, the exponentiation and division by variable (1/$x$) blocks are replaced with coupling weights consisting of floating gate memories operating in a subthreshold regime. Logarithmically encoded monomial values from the first array and log-encoded variable values are directly applied to the second array's gate and source voltage lines. As a result, division and exponentiation operations are executed by virtue of the exponential dependence of cell current on the difference of applied gate and source voltages. (d) Energy function and its partial derivatives and (e) crossbar circuit implementation for a more general polynomial (non-multi-linear). Note that a term with $x_4$ in the third power of the energy function is implemented with multi-state weights in the first crossbar and factor of 3 in the second crossbar.

```
p cnf 14  64
 5 10 13 0
3 -11 -12 0
7 -8 -12 0
0 -2 -7 0
-6 -7 13 0
2 -8 -11 0
5 7 -9 0
-6 -7 -11 0
-5 -7 8 0
0 -6 -12 0
2 4 9 0
1 -7 9 0
7 -8 -10 0
-1 5 12 0
-3 -6 13 0
0 1 -12 0
-1 6 10 0
1 5 -10 0
-1 -11 -13 0
2 4 -8 0
5 -6 -10 0
1 -6 13 0
9 10 13 0
0 5 10 0
2 5 11 0
-1 -2 5 0
0 -8 9 0
-8 -11 -12 0
10 -11 12 0
0 -2 13 0
0 -4 -6 0
-4 -5 -12 0
1 2 -11 0
2 -3 -13 0
11 12 -13 0
0 7 13 0
2 6 -8 0
-2 -6 7 0
-2 7 -12 0
0 -4 -10 0
-4 -6 -11 0
1 -2 -13 0
3 8 -11 0
-11 12 -13 0
-1 -4 8 0
2 -4 -11 0
-3 -8 10 0
-1 6 13 0
1 6 11 0
4 -10 11 0
0 -10 12 0
0 2 -11 0
9 11 -13 0
-1 -4 10 0
0 4 -10 0
4 9 -10 0
0 3 8 0
0 -1 -13 0
0 9 -12 0
1 11 -12 0
-4 8 11 0
2 -5 6 0
-1 5 -7 0
3 5 13 0
%
0
```

**Figure S11. 3-SAT instance solved in the experimental demo.**

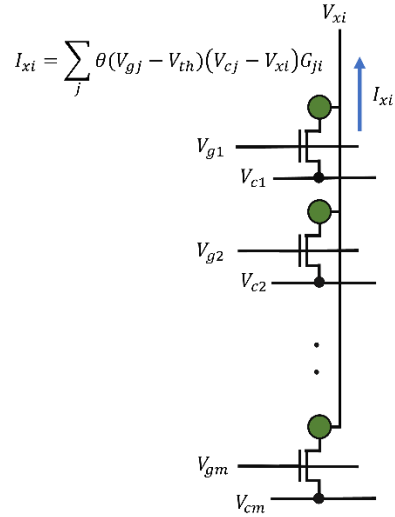

**Figure S12. Single bit line in a 1T1R memristor crossbar array.** The transistor in each cell acts like a switch, mathematically represented by the shifted Heaviside step function $\theta(V_{gj}-V_{th})$. When the switch is closed ($V_{gj}>=V_{th}$), a current proportional to the product of the voltage difference across the cell and the conductance of the memristor flows. When the switch is open ($V_{gj}<V_{th}$), zero current flows.

**<u>Supplementary Note 1</u>: Partial derivatives and difference quotients**

The gradient of a multi-variate continuous, differentiable function $H: R^N \to R$ is given by Equation 1 where $H_{xi}$ is the partial derivative of $H$ with respect to variable $x_i$ and is evaluated as shown in Equation 2.

$$\nabla H = [\frac{\partial H}{\partial x_1}, \frac{\partial H}{\partial x_2}, \ldots, \frac{\partial H}{\partial x_N}]^T = [H_{x1}, H_{x2}, \ldots, H_{xN}] \qquad (1)$$

$$H_{x1} = \lim_{h \to 0} \frac{H(x_1+h, x_2, \ldots, x_N) - H(x_1, x_2, \ldots, x_N)}{h} \qquad (2)$$

The partial derivative $H_{xi}$ tells the sign and the rate at which the function $H$ will change when variable $x_i$ is incremented or decremented by a certain amount. This guides gradient descent/ascent based algorithms to iteratively update the respective variables with correct update direction/magnitude to decrease/increase the value of $H$. Consider a degree one bi-variate polynomial function $H = 2x_1 - 3x_2$, where $H_{x1} = 2$ and $H_{x2} = -3$. This implies that when variable $x_1$ is increased the function $H$ increases with a rate of 2 whereas when variable $x_2$ is increased the function decreases with a rate of 3.

On the other hand, for discrete-valued functions like $H:\{a, a+h, \ldots, a+Kh\}^N \to R$ (where $\{a, a+h, \ldots, a+Kh\}$ is the discrete domain of the function with $a$, $K$ and $h$ being integers), the usual definitions of gradients and partial derivatives are not useful. For such functions, the analogue to derivative is the partial difference quotient (see Ref. 44) which is given by Equation 3. Note how at the limit $h \to 0$, the difference quotient becomes the partial derivative $H_{xi}$ in continuous space.

$$H_{x1} = \frac{\Delta H}{\Delta x_1} = \frac{H(x_1+h, x_2, \ldots, x_N) - H(x_1, x_2, \ldots, x_N)}{h} \qquad (3)$$

Now for a binary valued function $H:\{0,1\}^N \to R$, the domain of each variable contains *0* and *1*. So when $x_i=0$, its only valid move is to change to 1 and vice versa. The difference quotients of H for $x_1=0$ and $x_1=1$ are shown in Equations 4 and 5 (we have omitted the non-participating variables for simplicity). Note that they are equal to each other.

$$H_{x1}|_{x_1=0} = \frac{\Delta H}{\Delta x_1}|_{x_1=0} = \frac{H(x_1=1)-H(x_1=0)}{1-0} = H(x_1=1) - H(x_1=0) \quad (4)$$

$$H_{x1}|_{x_1=1} = \frac{\Delta H}{\Delta x_i}|_{x_1=1} = \frac{H(x_1=0)-H(x_1=1)}{0-1} = H(x_1=1) - H(x_1=0) \quad (5)$$

In the context of gradient descent inspired local search algorithms for solving combinatorial optimization problems that have such objective functions, the difference quotient provides information about the magnitude but not the sign/direction in which the function $H$ will change if a variable undergoes a valid move. What is more important for such algorithms is the net change in the function $H$ when a variable $x_i$ changes from *0* to *1* or from *1* to *0*, where both the magnitude and the sign of the change is used to decide whether to accept that move or not. This is simply obtained by taking the product of the difference quotient with the difference in the final and initial variable values and is shown in Equation 6.

$$\Delta H_{xi} = H_{xi} \Delta x_i \quad (6)$$

Throughout the remainder of the supplementary and main text, we shall refer to $\Delta H_{xi}$ in the context of binary valued functions H, as the "pseudo derivative of *H* w.r.t. $x_i$" or simply the "pseudo partial derivative". The pseudo derivative $H_{xi}$ is the net change in the function H when binary variable $x_i$ is flipped, i.e., when it changes its value to $\bar{x}_i$. The vector of pseudo derivatives of a multivariate function *H* with respect to all its variables is referred to as the "pseudo gradient".

**Supplementary Note 2: Gradient computation in high-order polynomials**

Any *K*-order discrete multi-variate polynomial in variables *{$x_1$, $x_2$, ..., $x_N$}* $\in X$, can be expressed as a summation of monomials (terms) - see Equation (7), where $t_j$ is the $j^{th}$ monomial and $t_j \in T$, the set of all monomials of H. Each of these monomials is a product of at least one variable (*H* might also include a constant offset term, but it is never relevant to computing derivatives, and is removed from *T*) and a real factor $a_j$. Further, let *E* be the set of all pairs *{$x_i, t_j$}* such that degree of $x_i$ in monomial $t_j$ is non-zero (see Equation (7)).

$$H = \sum_{j=1}^{M} t_j, \quad t_j = a_j \prod_{i \,\forall\, \{x_i,t_j\} \in E}^{N} x_i \quad (7)$$

For such a polynomial, the pseudo derivative with respect to variable $x_f$ (net change in the polynomial on flipping variable $x_f$, *i.e.,* changing value from "pre" state 0 to "post" state 1 or vice versa) is given by equation (8). The variable $x_f$ is factored out and the rest is a sum of partial monomials. The variable $x_f$ is factored out and the rest is a sum of partial monomials, with degree one less (in this case, removing $x_f$) than their respective original monomials.

$$\Delta H_{xf} = \sum_{j \,\forall\, \{x_f,t_j\} \in E} \Delta t_j = \sum_{j \,\forall\, \{x_f,t_j\} \in E} (t_j^{post} - t_j^{pre}) = \Delta x_f \sum_{j \,\forall\, \{x_f,t_j\} \in E} a_j \prod_{\substack{i \,\forall\, \{x_i,t_j\} \in E,\\ i \neq f}}^{N} x_i \quad (8)$$

Since the variables are binary, two cases arise — Case I: $x_f^{pre} = 1$ ($x_f^{post} = 0$), where initial value of $x_f$ is 1 and Case II: $x_f^{pre} = 0$ ($x_f^{post} = 1$), where initial value of $x_f$ is *0*.

Case I ($x_f^{pre} = 1$):

The pseudo partial derivative with respect to the now-zero-valued variable is computed as:

$$\Delta H_{xf} = \sum_{j \,\forall\, \{x_f,t_j\} \in E}(t_j^{post} - t_j^{pre}) = \sum_{j \,\forall\, \{x_f,t_j\} \in E} -t_j^{pre} = -\sum_{j \,\forall\, \{x_f,t_j\} \in E} a_j \prod_{i \,\forall\, \{x_i,t_j\} \in E}^{N} x_i \quad (9)$$

We define ancillary variable $z_j = \prod_{i \,\forall\, \{x_i,t_j\} \in E} x_i$ (local to monomial $t_j$ and consistent with the *z* introduced in Fig. 1e of main text) and equal to the product of the variables in that term. Observing that the product of several binary variables is one if and only if each multiplicand is one, and that the order of the term is equal to the number of multiplicands, the variables $z_j$ can also be computed as in Equation (10), where $K_j$ is the order of the term. This replaces a product with a linear transformation, which allows for its massively parallel in-memory computation. We refer to this variable as the "break monomial indicator" as it indicates if the monomial's value would change from non-zero to zero on flipping of any of its member variables.

$$z_j = \begin{cases} 1 & if \sum_{i \forall \{x_i,t_j\} \in E} x_i = K_j \\ 0 & otherwise \end{cases} \qquad (10)$$

We define $B$ as the $M \times N$ adjacency matrix of the graph $G$, where $B = (\hat{b}_1, \hat{b}_2, ..., \hat{b}_N)$ such that $\hat{b}_i = (b_{1i}, b_{2i}, ..., b_{Mi})^T$ and $b_{ji} = 1$ if $\{x_i, t_j\} \in E$ and $0$ otherwise. A vector-matrix multiplication between this adjacency matrix and the variables can then be used to compute the sum of variables in each term (see equation 11 below), where $\hat{\Sigma}_j^{mono} = (\Sigma_1^{mono}, \Sigma_2^{mono}, ..., \Sigma_M^{mono})$ is the vector representing the sum of variables in each monomial (referred to as the monomial variable sum in Fig. 1c) and computed in the forward pass (see Fig 1c of main text) and $\hat{x} = (x_1, x_2, ..., x_N)$ is the variable vector. The expression for $z_j$ can then be re-written as shown in Equation (12).

$$\hat{\Sigma}_j^{mono} = \hat{x} B^T \qquad (11)$$

$$z_j = \begin{cases} 1 & if \ \Sigma_j^{mono} = K_j \\ 0 & otherwise \end{cases} \qquad (12)$$

Substituting the break-monomial indicator in the pseudo derivative expression (of equation 9) and considering the monomial factors, one observes Equation (13) where $\Sigma_f^{break}$ is referred to as the "break variable sum" with respect to variable $x_f$. It is the number of monomials weighted by their respective factors that will become zero from non-zero when a one-valued variable $x_f$ is flipped to zero.

$$\Delta H_{xf} = -\Sigma_f^{break} = -\sum_{j=1}^M a_j z_j b_{jf} \qquad (13)$$

Case II ($x_f^{pre} = 0$):

The pseudo partial derivative with respect to the previously-zero-valued variable is computed as shown in Equation (14).

$$\Delta H_{xf} = \sum_{j \forall \{x_f,t_j\} \in E}(t_j^{post} - t_j^{pre}) = \sum_{j \forall \{x_f,t_j\} \in E} t_j^{post} = x_f^{post} \sum_{j \forall \{x_f,t_j\} \in E} a_j \prod_{\substack{i \forall \{x_i,t_j\} \in E, \\ i \neq f}}^N x_i \qquad (14)$$

Similarly, when $x_f^{pre} = 0$ ($x_f^{post} = 1$), the monomial values pre-variable-flipping ($t_j^{pre}$) are equal to zero. Since $x_f^{post} = 1$, it can be factored out and set to 1. Note that since $x_f^{pre} = 0$, the product in the right-most side of equation (14) is equal to one only when all but one variable ($x_f$) in the monomial $t_j$ have a value of 1. To evaluate this condition using linear transformation, we define a second ancillary variable $s_j$ (local to term $t_j$) (see equation (15)) which is 1 when the sum of variables in the monomial is exactly equal to one less than the order of that term ($K_j$) and 0 otherwise. We refer to this variable as the "make monomial indicator" (see Fig. 1d of main text) as it indicates if the monomial's value will change from zero to non-zero on flipping of one of its member variables. The pseudo derivative for the $x_f^{pre} = 0$ case, can then be succinctly written down as in Equation (16), where $\Sigma_f^{make}$ is the make variable sum with respect to variable $x_f$. It is the number of monomials weighted by their respective factors that will become non-zero from zero when a zero-valued variable $x_f$ is flipped to one.

$$s_j = \begin{cases} 1 & if \ \Sigma_j^{mono} = K_j - 1 \\ 0 & otherwise \end{cases} \qquad (15)$$

$$\Delta H_{xf} = \Sigma_f^{make} = \sum_{j \forall \{x_f,t_j\} \in E} a_j \prod_{\substack{i \forall \{x_i,t_j\} \in E, \\ i \neq f}}^N x_i = \sum_{j=1}^M a_j s_j b_{jf} \qquad (16)$$

Since the pseudo derivative with respect to one-valued variables is given by the break-variable sum (Equation 13) and that with respect to zero-valued variables is given by the make-variable sum (Equation 16), both can be combined as shown in Equation (17) to provide a generalized expression for the pseudo derivative of all the variables, that is equal to the difference between the make and break values.

$$\Delta \hat{H} = (1 - \hat{x}) \circ \hat{\Sigma}^{make} - \hat{x} \circ \hat{\Sigma}^{break} = ((\hat{a} \circ \hat{s})B) \circ (1 - \hat{x}) - ((\hat{a} \circ \hat{z})B) \circ \hat{x} \qquad (17)$$

**Supplementary Note 3: Converting CNF to high order polynomial.**

A CNF formula comprises of conjunction of clauses (from a set $C=\{c_1, c_2, .. , c_M\}$ of $M$ clauses) where a clause is a disjunction of literals (from a set $L=\{l_1, l_2,..., l_{2N-1}, l_{2N},\}$ s.t, $l_{2i-1} = x_i$ (positive literal of variable $x_i$), $l_{2i} = \bar{x}_i$ (negated literal of variable $x_i$) of 2N literals arising out of N variables) or in other words product of sums (AND of ORs). For example, there are two clauses in the CNF shown in Fig. 2a. The first clause has four literals and the second has two. One way of expressing the energy function of a generic CNF problem is shown in equation (18), where $w_j$ is the weight of the clause and $U_j$ is the indicator variable that tracks satisfaction of clause j. The merit function H then denotes the weighted sum of unsatisfied clauses and the objective is to minimize it.

$$H = \sum_{j=1}^{M} w_j U_j \; , \qquad U_j = \begin{cases} 0 \text{ if } c_j \text{ is satisfied} \\ 1 \text{ if } c_j \text{ is unsatisfied} \end{cases} \tag{18}$$

The variable $U_j$ of each clause can be expressed as a product of linear functions of its member variables as shown in Equation (19). For each of its member variable $x$, if its positive literal ($l_{2i-1}$) is present then the linear function is $(1-x)$, and if its negated literal ($l_{2i}$) is present then the linear function is $x$.

$$U_j = \prod_{i \, \forall \, \{l_{2i-1}, c_j\} \in E \text{ or } \atop \{l_{2i}, c_j\} \in E}^{K} g(x_i, c_j) \; , \quad \text{where } g(x_i, c_j) = \begin{cases} x_i \text{ if } \{l_{2i}, c_j\} \in E \\ 1 - x_i \text{ if } \{l_{2i-1}, c_j\} \in E \end{cases} \tag{19}$$

Take for example the CNF considered in equation (20), the energy function can then be expanded as in equations 21-22:

$$CNF = (\bar{x}_1||x_2||\bar{x}_3) \; \& \; (\bar{x}_2||x_3||x_4) \; \& \; (x_3||x_1||\bar{x}_4) \tag{20}$$

$$H = x_1(1-x_2)x_3 + x_2(1-x_3)(1-x_4) + (1-x_3)(1-x_1)x_4 \tag{21}$$

$$H = x_1 x_3 - x_1 x_2 x_3 + x_2 - x_2 x_3 - x_2 x_4 + x_2 x_3 x_4 + x_4 - x_3 x_4 - x_1 x_4 + x_1 x_3 x_4 \tag{22}$$

Note how three clauses result in 10 terms in the resulting higher order polynomial. Such a conversion can lead to $O(2^k)$ many terms in the polynomial arising out of each clause (with clause size $k$) in the original CNF in the worst case.

**Supplementary Note 4: Gradient computation in Conjunctive-Normal Form**

Any Constraint Satisfaction Problem (CSP) in CNF can be represented by a bipartite graph $G \in \{L,C,E\}$, where one set of vertices $L$ denote the literals $\{l_1, l_2,..., l_{2N-1}, l_{2N},\}$ s.t, $l_{2i-1} = x_i$ (positive literal of variable $x_i$), $l_{2i} = \bar{x}_i$ (negated literal of variable $x_i$), the other set $C$ the clauses $\{c_1, c_2, .. , c_M\}$ and $E$ the set of edges defined by $\{l_i, c_j\}$ such that literal $l_i$ is a member of clause $c_j$. As such, there are $N$ variables, $2N$ literals and $M$ clauses in the CNF. Let **B** denote the forward binary $M \times 2N$ adjacency matrix of $G$, where $\mathbf{B} = (\hat{b}_1, \hat{b}_2, ..., \hat{b}_{2N})$ such that $\hat{b}_i = (b_{1i}, b_{2i}, ..., b_{Mi})^T$ and $b_{ji} = 1$ if $\{l_i, c_j\} \in E$ and $0$ otherwise.

The objective function in such CNF-based problems is the total number of unsatisfied clauses weighed by the clause weights $w_j$, where a clause is satisfied if the OR of its member literals is true and unsatisfied otherwise (see equation 23). The goal of the optimization problem is then to minimize $H$. Therefore, the definition of pseudo derivatives here is the net change in the sum of weights of unsatisfied clauses when a variable is flipped. A popular score metric known as the gain is used in optimization algorithms designed for solving CSPs that are present in CNF. It is the net increase in the number of satisfied clauses weighed by their respective clause weights and can be expressed as the negative of the CNF pseudo derivative (see Equation 24).

$$H = \sum_{j=1}^{M} w_j U_j \; , \qquad U_j = \begin{cases} 0 \text{ if } c_j \text{ is satisfied} \\ 1 \text{ if } c_j \text{ is unsatisfied} \end{cases} \tag{23}$$

In a CNF, when a literal is flipped not only do the clauses that contain that literal change but also the ones that contain its complement. Therefore, one needs to track the change caused by both literals associated with a given variable. The CNF gain when a variable $x_f$ is flipped can then be expressed as the sum of gains with respect to its two literals $l_{2f-1}$ and $l_{2f}$.

$$gain_{x_f} = gain_{l_{2f-1}} + gain_{l_{2f}} = -\sum_{j \, \forall \, \{l_{2f-1}, c_j\} \in E} w_j \Delta U_j - \sum_{j \, \forall \, \{l_{2f}, c_j\} \in E} w_j \Delta U_j \tag{24}$$

Note that out of the two literals, one of them is true (one) and the other is false (zero) in any given configuration, so we evaluate the gradient contributions of the two separately. We define $pl(x_f)$ as the literal of $x_f$ that is currently 1 and $nl(x_f)$ as the literal of $x_f$ that is 0. Equation (24) can then be re-written as:

$$gain_{x_f} = gain_{nl(x_f)} + gain_{pl(x_f)} \qquad (25)$$

Interestingly, $gain_{pl(x_f)}$ is always a negative value and its absolute value is popularly known as the break-value (break$_{xf}$, analogous to but not the same as break-values that we defined in context of polynomial pseudo derivatives). This is because a clause (in CNF, a disjunction) is satisfied when any one of its literals is one. So when a literal that is currently 1 is flipped to 0, it changes the clause status from satisfied to unsatisfied or remains the same. Therefore, break-value depicts the number of clauses that changed from satisfied to unsatisfied. Similarly, $gain_{nl(x_f)}$ is always a positive value and is known as the make-value (make$_{xf}$, analogous to but not the same as make-values that we defined in context of polynomial pseudo derivatives). It depicts the number of clauses that changed from unsatisfied to satisfied. Before going ahead with the gain calculation, the linear transformation in the forward-step (a matrix-vector multiplication between the literal-vector and the adjacency matrix) that computes the sum of the literal-values in each clause is defined in Equation 26.

$$\widehat{\Sigma}^{clause} = \hat{l}\,B^T \qquad (26)$$

Case I (for the zero-valued literal, $nl(x_f)$):

The gain with respect to the zero-valued literal $nl(x_f)$ can be written as shown in Equation (27).

$$gain_{nl(x_f)} = -\sum_{j\,\forall\,\{nl(x_f),c_j\}\in E} w_j \Delta U_j \;=\; -\sum_{j\,\forall\,\{nl(x_f),c_j\}\in E} w_j (U_j^{post} - U_j^{pre}) \qquad (27)$$

The variable $U_j$ can be substituted with the Kronecker-delta function as in equation (28). $U_j$ equates to one when the sum of its member variables is zero (clause is unsatisfied) and equates to zero when the sum of its member variables is non-zero (clause is satisfied).

$$U_j = \delta[\sum_{i\,\forall\,\{l_i,c_j\}\in E} l_i] = \delta[\Sigma_j^{clause}] \qquad (28)$$

Since, equation (27) evaluates the CNF gain for literals that are currently 0, all clauses containing that literal would become satisfied post-flipping i.e., $U_j^{post}=0$ (see Equation 29).

$$gain_{nl(x_f)} = -\sum_{j\,\forall\,\{nl(x_f),c_j\}\in E} w_j (0 - U_j^{pre}) \;=\; \sum_{j\,\forall\,\{nl(x_f),c_j\}\in E} w_j \delta[\Sigma_j^{clause}]) \qquad (29)$$

A clause-local variable $s_j$ (also referred to as the make-clause indicator in Fig. S5d) as defined in equation (30) can then be substituted in equation (29) to yield the following expression for the gain as shown in equation (31). Here $\sum_j w_j b_{j,nl(x_f)} s_j$ is the output that is computed in the backward make pass VMM (see Fig. S5d) and is referred to as the make-literal sum (see Equation 32) of literal $nl(x_f)$. Since $nl(x_f)$ is the zero-valued literal, its make-literal sum is also the make-value of variable $x_f$. The make-value of a variable is therefore the make literal sum (backward make pass output) of its zero-valued literal. To generalize the expression in Equation 31 and write it without exact information about the zero-valued literal, we write the make-value of $x_f$ as the sum of make-literal sums scaled by the inverted literal value of both its literals as shown in Equation 33.

$$s_j = \begin{cases} 1 & if\ \Sigma_j^{clause} = 0 \\ 0 & otherwise \end{cases} \qquad (30)$$

$$gain_{nl(x_f)} = \sum_{j\,\forall\,\{nl(x_f),c_j\}\in E} w_j s_j \;=\; \sum_{j=1}^{M} w_j b_{j,nl(x_f)} s_j \qquad (31)$$

$$\Sigma_{nl(x_f)}^{make} = \sum_{j=1}^{M} w_j b_{j,nl(x_f)} s_j \qquad (32)$$

$$gain_{nl(x_f)} = (1-l_{2f-1})\Sigma_{l_{2f-1}}^{make} + (1-l_{2f})\Sigma_{l_{2f}}^{make} \;=\; (1-l_{2f-1})\sum_{j=1}^{M} w_j b_{j,2f-1} s_j + (1-l_{2f})\sum_{j=1}^{M} w_j b_{j,2f} s_j \quad (33)$$

Case II (for the one-valued literal, $pl(x_f)$):

The gain with respect to the one-valued literal $pl(x_f)$ can be written as shown in Equation (34).

$$gain_{pl(x_f)} = -\sum_{j \, \forall \, \{pl(x_f),c_j\} \in E} w_j \left( U_j^{post} - U_j^{pre} \right) = -\sum_{j \, \forall \, \{pl(x_f),c_j\} \in E} w_j \left( U_j^{post} - 0 \right) \qquad (34)$$

Since we are considering the gain for $pl(x_f)$ (the literal that is currently one), $U_j^{pre}$ can be set to 0 in equation (34), since clause $j$ is satisfied. Subsequently, $U_j^{post}$ can be replaced with $\delta[\sum_j^{clause}\text{-}1]$, where $\sum_j^{clause}$ denotes the sum of literals pre-flipping and the '1' is subtracted from it to reflect the literal $pl(x_f)$ changing from 1 to 0.

$$gain_{pl(x_f)} = -\sum_{j \, \forall \, \{pl(x_f),c_j\} \in E} w_j \delta[\sum_j^{clause} - 1] \qquad (35)$$

Additionally, we define another clause-local variable $z_j$ known as the break-clause indicator (see Fig. S5e) as shown in equation (36).

$$z_j = \begin{cases} 1 & if \sum_j^{clause} - 1 = 0 \\ 0 & otherwise \end{cases} = \begin{cases} 1 & if \sum_j^{clause} = 1 \\ 0 & otherwise \end{cases} \qquad (36)$$

Substituting this new variable in the gain expression, the final gain can then be written as shown in equation (37), where $\sum_{j=1}^{M} w_j b_{j,pl(x_f)} z_j$ is the output that is computed in the backward break pass VMM and is known as the break-literal sum of literal $pl(x_f)$ denoted by $\sum_{pl(x_f)}^{break}$ (see Fig. S5e).

$$gain_{pl(x_f)} = -\sum_{j \, \forall \, \{pl(x_f),c_j\} \in E} w_j z_j = -\sum_{j=1}^{M} w_j b_{j,pl(x_f)} z_j = -\sum_{pl(x_f)}^{break} \qquad (37)$$

Since $pl(x_f)$ is the one-valued literal, its break-literal sum is also the break-value of variable $x_f$. The break-value of a variable is therefore the break literal sum (backward break pass output) of its one-valued literal. To generalize the expression in Equation (37) and write it without exact information about the one-valued literal, we write the gain of $pl(x_f)$ as the sum of break-literal sums scaled by the non-inverted literal value of both literals of $x_f$ as shown in Equation (38). The absolute value of the expression in Equation (38) is the break-value of variable $x_f$.

$$gain_{pl(x_f)} = -l_{2f-1}\sum_{l_{2f-1}}^{break} - l_{2f}\sum_{l_{2f}}^{break} = -l_{2f-1}\sum_{j=1}^{M} w_j b_{j,2f-1} z_j - l_{2f} \sum_{j=1}^{M} w_j b_{j,2f} z_j \qquad (38)$$

The net gain of the variable is the sum of gains associated with its positive and negated literals or in other words the difference between make and break-values (see Equation 25) and therefore it can be written as shown in Equations 39-40. The clause-weight vector $\widehat{w}$ is unitary for k-SAT problems.

$$gain_{x_f} = \{(1 - l_{2f-1})\sum_{l_{2f-1}}^{make} + (1 - l_{2f})\sum_{l_{2f}}^{make}\} - \{l_{2f-1}\sum_{l_{2f-1}}^{break} + l_{2f}\sum_{l_{2f}}^{break}\} \qquad (39)$$

$$gain_{x_f} = \{(1 - l_{2f-1})(\widehat{w} \circ \hat{s}) \cdot \hat{b}_{2f-1} + (1 - l_{2f})(\widehat{w} \circ \hat{s}) \cdot \hat{b}_{2f}\} - \{l_{2f-1}(\widehat{w} \circ \hat{z}) \cdot \hat{b}_{2f-1} + l_{2f}(\widehat{w} \circ \hat{z}) \cdot \hat{b}_{2f}\} \qquad (40)$$

**Supplementary Note 5: Single-step backward pass with three-terminal crossbar arrays.**

When passive cross point devices (like 0T1R memristors) are replaced with three-terminal devices (like 1T1R memristor bit-cells), the backward step can be executed in the same crossbar array in a single cycle, instead of using two separate arrays like shown in Fig. S3c. To demonstrate this, consider the gradient computation in polynomial forms and CNF separately as described below.

Case A (Polynomials):

Using Equation 17, one can write down the gradient with respect to a single variable $x_f$ as follows (we assume all monomial factors to be one for simplicity):

$$\Delta H_{x_f} = (1 - x_f)\sum_f^{make} - x_f \sum_f^{break} = \hat{s} \cdot \hat{b}_f (1 - x_f) - \hat{z} \cdot \hat{b}_f x_f = (-\hat{z} x_f + \hat{s} - \hat{s} x_f) \cdot \hat{b}_f \qquad (41)$$

Note that we can rewrite $\hat{s} = \hat{s} \circ \hat{s}$ (as elementwise product with itself) without loss of generality since it is a binary-valued vector. Also, based on definitions of $\hat{s}$ and $\hat{z}$ (in equation 12 and 15), one can see that both vectors cannot have their entries at the same position to be equal to one. This means $s_j \times z_j$ is always equal to zero. This allows us to re-write equation (41) as shown below:

$$\Delta H_{xf} = (\hat{s} \circ \hat{z} - \hat{z}\, x_f + \hat{s} \circ \hat{s} - \hat{s}\, x_f).\hat{b}_f \;=\; \left((\hat{s} + \hat{z}) \circ (\hat{s} - x_f)\right).\hat{b}_f \tag{42}$$

The same equation can then be expanded using a summation as seen below:

$$\Delta H(x_f) = \sum_{j=1}^{M}(s_j + z_j)(s_j - x_f)\, b_{fj} \tag{43}$$

The expression inside the summation of Equation (43) denotes a triple product and is identical to the expression (Equation 44) for current flowing in a bit line of a 1T1R crossbar array (see Fig. S12).

$$I_f = \sum_{j=1}^{M} \theta(V_{gj} - V_{th})(V_{Cj} - V_{Xf})\, G_{jf} \tag{44}$$

Since the signal $s_j+z_j$ is binary ("+" denoting arithmetic addition here, not logical disjunction, as $s_j$ and $z_j$ cannot be unity at the same given time) it acts as the gating signal $gate_j$. Depending on whether $gate_j$ is 0 or 1, the voltage $V_{gj}$ applied to the gate of the transistor in Fig. S12 is chosen to turn the switch off or on respectively ($\theta(V_{gj} - V_{th}) = 1$ when $V_{gj} \geq V_{th}$ and $= 0$ when $V_{gj} < V_{th}$). Voltage $V_{Cj}$ proportional to $s_j$ ($=s_jV_0$) is applied to the other terminal of the transistor from the monomial side (along the rows) and $V_{xi}$ proportional to variable value $x_f$ ($=x_fV_0$) is applied along the column lines from the variable side. The normalization factor between Equations (43) and (44) is the unit current via on-state coupling weight, $I_0 \equiv V_0 G_{on}$.

Case B (CNFs):

One can compactly re-write Equation (40) as shown in Equation (45) (we assume all clause weights to be one).

$$gain_{x_f} = \sum_{k=0}^{1}\big(\hat{s}(1 - l_{2f-k}) - \hat{z}\, l_{2f-k}\big).\hat{b}_{2f-k} \tag{45}$$

After doing manipulations like what was introduced in Equation (42) (in Case A), we get Equation (46).

$$gain_{x_f} = \sum_{k=0}^{1}(\hat{s} - \hat{s}\, l_{2f-k} - \hat{z}\, l_{2f-k}).\hat{b}_{2f-k} \;=\; \sum_{k=0}^{1}(\hat{s} \circ \hat{z} + \hat{s} \circ \hat{s} - \hat{s}\, l_{2f-k} - \hat{z}\, l_{2f-k}).\hat{b}_{2f-k} \;=\; \sum_{k=0}^{1}(\hat{s} + \hat{z}) \circ (\hat{s} - l_{2f-k}).\hat{b}_{2f-k} \tag{46}$$

Re-writing the same in the summation form we get Equations 47-48.

$$gain_{x_f} = \sum_{j=1}^{M}\sum_{k=0}^{1}(s_j + z_j)(s_j - l_{2f-k})\, b_{j,2f-k} \tag{47}$$

$$gain_{x_f} = \sum_{j=1}^{M}(s_j + z_j)(s_j - x_f)\, b_{j,2f-1} + \sum_{j=1}^{M}(s_j + z_j)(s_j - \bar{x}_f)\, b_{j,2f} \tag{48}$$

$$I_{xf}^{total} = I_{xf} + I'_{xf} = \sum_{j=1}^{M} \theta(V_{gj} - V_{th})(V_{cj} - V_{xf})\, G_{j,2f-1} + \sum_{j=1}^{M} \theta(V_{gj} - V_{th})(V_{cj} - V'_{xf})\, G_{j,2f} \tag{49}$$

Like Case A, the two inner summations expressed in equation (47-48) can be mapped to the current flowing through a pair of bit-lines (corresponding to the two literals of a variable) of a 1T1R crossbar array that is summed at the periphery as shown in Equation (49) (see Fig. S8). Depending on whether the binary-valued signal $gate_j = s_j + z_j$ is 0 or 1, the transistor gate voltage $V_{gj}$ ((see Fig. S8c)) is set such that it turns the transistors in that row off or on respectively ($\theta(V_{gj} - V_{th}) = 1$ when $V_{gj} \geq V_{th}$ and $= 0$ when $V_{gj} < V_{th}$). The other set of voltages $V_{cj}$ applied from the clause side (along the rows in Fig. S8) is proportional to the make clause indicator signal $s_j$ ($=s_jV_0$), while the voltages $V_{xf}$ and $V'_{xf}$ applied along the columns (in Fig. S8) from the literal side are proportional to the literal values $x_f$ ($=x_fV_0$) and $\bar{x}_f$ ($=\bar{x}_fV_0$) respectively. The $b_{j,2f-k}$ values are mapped to the conductance $G_{j,2f-k}$ of the crosspoint devices. The normalization factor between equations 48 and 49 is the unit current via on-state coupling weight, $I_0 \equiv V_0 G_{on}$.